\documentclass[12pt]{iopart}
\usepackage{amsthm,graphicx,latexsym,amssymb,esint,color, mathrsfs} 
\usepackage{setstack}
\usepackage{iopams}

\graphicspath{{pdf/}}

\makeatletter

\@addtoreset{equation}{section}
\makeatother

\newcommand{\nn}{\nonumber}

\newcommand{\half}{\frac{1}{2}}
\renewcommand{\Re}{{\rm Re}}
\renewcommand{\Im}{{\rm Im}}
\newcommand{\wh}{\widehat}
\newcommand{\wt}{\widetilde}
\newcommand{\vphi}{\varphi}
\newcommand{\Ub}{\overline{U}}
\newcommand{\Gammab}{\overline{\Gamma}}
\newcommand{\myid}{{\bf 1}}
\newcommand{\even}{{\rm ev}}
\newcommand{\odd}{{\rm odd}}
\newcommand{\mod}{\ \mathrm{mod}\;}

\newcommand{\kker}{\mathcal{K}}

\newcommand{\corr}[1]{#1}

\newcommand{\secref}[1]{Sec.~\ref{sec:#1}}
\newcommand{\figref}[1]{Fig.~\ref{fig:#1}}

\newenvironment{myfig}{\begin{figure}\begin{center}}{\end{center}\end{figure}}

\begin{document}

\title{Non-Linear Integral Equations for the $\mathrm{SL}(2,\mathbb{R})/\mathrm{U}(1)$
  black hole sigma model}

\author{Constantin Candu$^a$ and Yacine Ikhlef\,$^b$}
\address{$^a$ Institut f. Theoretische Physik, ETH Z\"urich, CH-8093 Z\"urich, Switzerland}
\address{$^b$ Universit\'e Paris 6, CNRS UMR 7589, LPTHE, 75252 Paris Cedex, France}
\ead{canduc@ipt.phys.ethz.ch, ikhlef@lpthe.jussieu.fr}

\begin{abstract}
  It was previously established that the critical staggered XXZ spin chain provides a lattice regularization of the black hole CFT. We reconsider the continuum limit of this spin chain with the exact method of non-linear integral equations (NLIEs), paying particular attention to the effects of a singular integration kernel. With the help of the NLIEs, we rederive the continuous
  black hole spectrum, but also numerically match the density of states of the spin chain with that of the CFT, which is a new result.
	Finally, we briefly discuss the integrable structure of the black hole CFT and the identification of its massive integrable perturbation on the lattice.
\end{abstract}


\tableofcontents
\title[NLIE for the black hole sigma model]{}



\section{Introduction}


The relationship between integrable spin chains and integrable quantum field theories has been a long and fruitful one.
In this respect,  the sine-Gordon (SG) model is an illustrative example.
Undoubtedly, its large volume physics can be entirely understood without any reference to a spin chain by means of the asymptotic Bethe ansatz, which is based on the exact (anti)soliton $S$-matrix of Zamolodchikov \& Zamolodchikov \cite{Zamolodchikov:1978xm}.
And although the vacuum energy in finite size can be computed from the IR data via the thermodynamic Bethe ansatz \cite{Yang:1968rm, Zamolodchikov:1989cf}, excited states are inaccessible this way.\footnote{The generalization in \cite{Balog:2003xd} of the SG TBA to multiparticle states uses the DDV equation derived from the lattice and, hence, it is not based on the IR data only.}
A different approach is based on the observation that the XXZ spin chain provides a lattice discretization for the SG model
in the sense that low energy excitations of the former scatter with the $S$-matrix of the latter \corr{\cite{Faddeev-Takh:1981}}.
One can then solve the XXZ spectrum problem in finite size by elementary means \cite{Klummpe:1991vs, Destri:1992qk}, take the continuum limit  and produce a single  non-linear integral equation (NLIE) solving the entire finite size spectrum problem for the SG quantum field theory  \cite{Destri:1997yz, Feverati:1998dt, Feverati:1998uz}.
The ultimate check of this NLIE is that it produces in the IR the required asymptotic multiparticle spectrum and in the UV the expected spectrum of a free massless compact boson.
Similar results were subsequently derived  for the affine Toda  theories with imaginary coupling  and  their unitary restrictions \cite{Hollowood:1992sy}, equivalent to massive perturbations of rational conformal field theories, from the analysis of integrable spin chains of higher rank quantum groups \cite{ZinnJustin:1997at}.

In contrast, our current understanding of integrable massive perturbations of
 non-rational CFTs, starting from first principles and including excited states, is limited to the IR region only, the only notable exception being the sinh-Gordon (ShG) model  \cite{Teschner:2007ng}.
To understand  why this is uncomforting,
take for instance the real affine-Toda theories.
These are characterized by a  solitonless massive spectrum and diagonal scattering \cite{Arinshtein:1979pb, Braden:1989bu}.
Naively, it seems rather surprising that a theory defined by such simple IR data can develop in the UV all the complicated features characteristic of a non-rational CFT, i.e.\ continuous spectrum, non normalizable vacuum, etc.
For the ShG model the mechanism of this process was understood in \cite{Teschner:2007ng}, building on earlier results from \cite{Zamolodchikov:2000kt}.
It is worth noticing that, again, the main tool was a NLIE for the finite size spectrum derived from a (tailor made) lattice ShG discretization \cite{Bytsko:2006ut}.
The remarkable observation of \cite{Teschner:2007ng, Bytsko:2009mg} was that the NLIE could encode both the scattering data in the IR and such fine non-rational CFT structures as reflection amplitudes in the UV.
Clearly, it would be nice to have more examples of perturbed non-rational CFTs for which one can study the evolution from IR to UV explicitly  by means of NLIEs derived from a lattice discretization.

Motivated by these considerations, we study  the integrable structure of  the  SL$(2,\mathbb{R})_k$/U(1) Euclidean black hole sigma model CFT\footnote{The black hole CFT is equivalent to the sine-Liouville model via the strong-weak coupling duality of Fateev, Zamolodchikov \& Zamolodchikov, see \cite{Hikida:2008pe} for a proof.}
defined by the (one loop) metric \cite{Witten:1991yr}
\begin{equation}
ds^2 = \frac{k}{2}(d\rho^2 + \tanh^2\rho\, d\varphi^2)\,,
\label{eq:cig_metr}
\end{equation}
starting from its lattice discretization as a staggered XXZ spin chain.
For this chain the emergence of a continuous spectrum in the continuum limit was first noticed in \cite{Jacobsen:2005xz},
further studied in \cite{Ikhlef:2008zz} and finally identified with the black hole spectrum in \cite{Ikhlef:2011ay}.
Our main result is a set of two NLIE for the black hole CFT, which we derive from the lattice.
The NLIEs  reproduce the conformal dimensions of all primary states in the  continuous component of the black hole spectrum \cite{Hanany:2002ev}, but also allow
to compute the eigenvalues of all mutually commuting local conserved charges of the CFT  on these primary states.
In this sense, our NLIEs characterize the quantum integrable structure of the black hole CFT, which is related to the quantization of the second Poisson structure of the non-linear Schr\"odinger hierarchy \cite{Schiff:1992tv}.
Coming back to our initial discussion of IR to UV flows, one expects that there is an integrable massive deformation of the black hole CFT which preserves this integrable structure and which can be realized straightforwardly on the lattice following the standard recipe of \cite{Destri:1987ug, Reshetikhin:1993wm}.
We shall discuss this point further in the concluding section.

Our NLIEs have some unusual features, some of which were expected \cite{Jacobsen:2005xz}. Firstly, the integral kernels defining them are ``singular'', i.e.\  do not decay at infinity.
Closely related to this fact is the appearance of unusual source terms and non-monotonic counting functions (even in the absence of holes).
To make sure that the NLIEs make sense, we have performed a very non-trivial check on them by matching numerically the density of states in the spin chain with the density of states in
the CFT.
More precisely, one can  imagine the target space of the black hole CFT as a semi-infinite cigar degenerating into a cylinder of radius $\sqrt{k/2}$ at asymptotic infinity.
Now if the continuous black hole spectrum is regularized by cutting off the infinite tail of the cigar with a Liouville wall as in \cite{Hanany:2002ev}, then we find that the resulting density of states agrees precisely with the
 density of spin chain states for which the discrete quantum numbers are kept fixed in the continuum limit.
Equivalently, the difference between the reflection amplitude at the tip of the cigar and off  the Liouville wall determines (a subleading term in)  the
asymptotic of the NLIE in the region where the Bethe roots condense.

The paper is structured as follows. In Sec.~\ref{sec:6v} we recall the staggered XXZ spin chain discretization of the black hole CFT, its Bethe ansatz solution and specify the class of excited states to which we restrict our subsequent analysis.
In Sec.~\ref{sec:NLIE1} we derive the NLIEs for the finite size spin chain spectrum and then take their continuum limit in Sec.~\ref{sec:NLIE2}.
We then put the latter NLIEs to work in Sec.~\ref{sec:spec}, where we compute the conformal spectrum of the spin chain and match it with the spectrum of the black hole.
We also give an integral representation for the generating functions of the local integrals of motion of the CFT.
Finally, in Sec.~\ref{sec:dens} we explain the numerical algorithm used to compute the density of states in the spin chain and compare the results to CFT predictions.
There is also a short appendix collecting some of the more technical calculations.

\section{Staggered six-vertex model}\label{sec:6v}
 

\subsection{Transfer matrix and conserved quantities}


\begin{myfig}
  \includegraphics[scale=1.2]{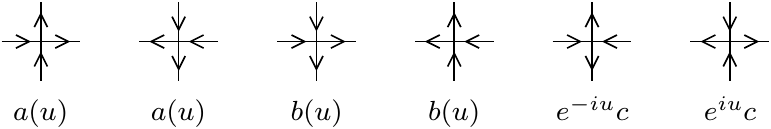}
  \caption{The configurations and Boltzmann weights of the six-vertex model.}
  \label{fig:6V}
\end{myfig}

\begin{myfig}
  \includegraphics[scale=1.2]{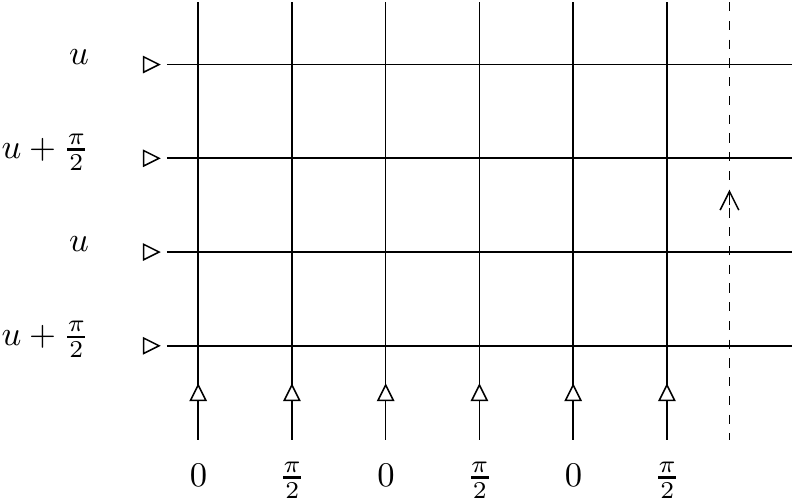}
  \caption{The staggered spectral parameters on the square lattice
    of width $2L$ sites.
    The dotted line represents the twisted periodic boundary conditions.}
  \label{fig:stag}
\end{myfig}

We consider the six-vertex (6V) model on the square lattice,
with the Boltzmann weights given by the $R$-matrix (see~\figref{6V}):
\begin{equation*}
  R(u) = \frac{1}{a(u)}\left(\begin{array}{cccc}
    a(u) & 0 & 0 & 0 \\
    0 & b(u)  & e^{-iu}c  & 0 \\
    0 &  e^{iu}c  & b(u) & 0 \\
    0 & 0 & 0 & a(u)
  \end{array}\right)
	\ ,\quad
	\begin{array}{cll}
	a(u)&=&\sin(\gamma-u)\,,\\
	b(u)&=&\sin u\,,\\
	c &=& \sin\gamma\,,
	\end{array}
\end{equation*}
where $u$ is the spectral parameter, and $\gamma$ defines the Baxter's
``anisotropy parameter'' $\Delta= (a^2+b^2-c^2)/2ab=-\cos \gamma$.
In this paper, we consider the regime:
\begin{equation} \label{eq:regime}
  0< \gamma < \frac{\pi}{2}
\,.
\end{equation}
The additional exponentials $e^{\pm iu}$ appearing in the off-diagonal terms of the $R$-matrix can be removed by a $\mathrm{U}(1)$-gauge transformation; keeping them has the advantage of making the $R$-matrix $\pi$-periodic.

Spectral parameters are carried by the
lines of the lattice, and the weights for a vertex with
spectral parameters $u$ and $v$ are given by $R(u-v)$.
We introduce a staggering of the horizontal and vertical
spectral parameters, as shown in~\figref{stag}.
For a row of $2L$ sites, the one-row transfer matrix with twisted periodic
boundary conditions is:
\begin{equation*}
  \fl \qquad  t(u) = {\rm Tr}_0 \, \left[ \exp(i\vphi \sigma^z_0)
    R_{0,2L} \left(u-\frac{\pi}{2} \right) R_{0,2L-1}(u)
    \dots
    R_{02} \left(u-\frac{\pi}{2} \right) R_{01}(u)
    \right] \,,
\end{equation*}
where we take $-\pi<\vphi<\pi$ and, for simplicity, restrict to $L$ even.
\begin{myfig}
  \includegraphics[scale=0.9]{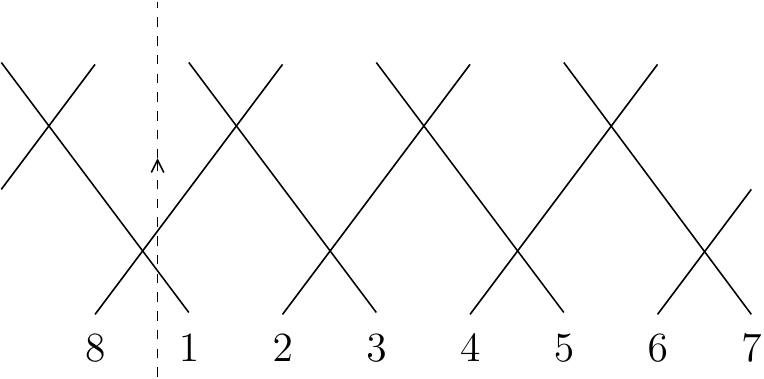}
  \caption{
    The action of the quasi-shift operator $\wt\tau$ (\ref{eq:def-taut}\,--\,\ref{eq:light-cone})
    for $2L=8$ sites with periodic boundary conditions.
    At each vertex, sits an $\check{R}(\pi/2)$ matrix acting in the vertical direction.
    The dotted line represents the twist.
  }
  \label{fig:light-cone}
\end{myfig}
The quantum Hamiltonian is defined by
\begin{equation} \label{eq:def-H}
  \fl \qquad H:= \half \sin 2\gamma \left[
    t^{-1}(0) \frac{dt}{du}(0) + t^{-1}(\pi/2) \frac{dt}{du}(\pi/2)
    \right] \,,
\end{equation}
and has the explicit form
\begin{eqnarray}
  \fl \qquad H &=& \sum_{j=1}^{2L} \left[
    -\half \boldsymbol{\sigma}_j \cdot\boldsymbol{\sigma}_{j+2}
    + \sin^2 \gamma \ \sigma_j^z \sigma_{j+1}^z
    \corr{-} \frac{i}{2} \sin\gamma \ (\sigma_{j-1}^z - \sigma_{j+2}^z)
    (\sigma_j^x \sigma_{j+1}^x + \sigma_j^y \sigma_{j+1}^y)
    \right] \nn \\
		\fl && \corr{+L\cos 2\gamma}\,, \label{eq:ham}
\end{eqnarray}
where $\sigma^a_j$ are the Pauli matrices at site $j$. In terms of the conserved $\mathrm{U}(1)$ charge
\begin{equation*}
S^z=\frac{1}{2}\sum_{j=1}^{2L}\sigma_j^z 
\end{equation*}
the twisted periodic boundary conditions for the local spin operators are given by
\begin{equation*}
  \sigma_{2L+j}^a= e^{2i\varphi S^z} \sigma_{j}^a e^{-2i\varphi S^z}\,.
\end{equation*}
The momentum operator is determined by the two-row transfer matrix at $u=0$ by
\begin{equation}\label{eq:mom}
   e^{iP}:=t(\pi/2)t(0) = \exp[i\vphi(\sigma_1^z+\sigma_2^z)] \ \tau \,,\qquad  e^{iPL} = e^{2i\varphi S^z}\,,
\end{equation}
where $\tau$ is the two-site translation operator.
Similarly, we define the {\it quasi-shift} operator
\begin{equation} \label{eq:def-taut}
  \wt\tau := t(\pi/2) t^{-1}(0) \,,
\end{equation}
which will play an important role in the following.
A little algebra shows that $\wt\tau$ has the form of
a diagonal-to-diagonal (or ``light-cone'') transfer matrix,
as depicted in~\figref{light-cone}:
\begin{equation} \label{eq:light-cone}
  \wt\tau = \left[\prod_{j=1}^L \check{R}_{2j,2j+1}(\pi/2) \right]
  \times e^{i\vphi\sigma_1^z}
  \times \left[\prod_{j=1}^L \check{R}_{2j-1,2j}(\pi/2) \right]
  \times e^{-i\vphi\sigma_1^z} \,,
\end{equation}
where $\check{R}_{12}(u):=P_{12} R_{12}(u)$, and $P_{12}$ is
the permutation operator. Let us notice  the relation $R_{ij}(\pi/2)R_{ji}(\pi/2)=I$,
which is useful for deriving
eqs.~(\ref{eq:mom}, \ref{eq:light-cone}).

As usual, higher order (local) conserved quantities can be generated by expanding the logarithms of the transfer matrices $t(u)$ and $t(u+\pi/2)$ around $u=0$.


\subsection{Bethe-Ansatz solution}


The eigenvalues of $t(u)$ take the well known form (see for instance \cite{IKB})
%
%
%
\begin{eqnarray}
  \fl\qquad \Lambda(u) &=& e^{i\vphi} \frac{\mathcal{Q}(u-\pi+\gamma)}{\mathcal{Q}(u)}
  + e^{-i\vphi} \left[\frac{\sin 2u}{\sin 2(u-\gamma)}\right]^L
  \frac{\mathcal{Q}(u+\pi-\gamma)}{\mathcal{Q}(u)}
       \,, \label{eq:Lambda}
\end{eqnarray}
where $\mathcal{Q}(u)=\prod_{j=1}^r \sinh\half[2iu-i\gamma+\lambda_j]$
and the parameters $\{\lambda_j\}_{j=1,\dots,r}$ are Bethe roots, which must be mutually distinct mod $2i\pi$ and solve the Bethe Ansatz Equations (BAE)
\begin{equation}
 \left[\frac{\sinh(\lambda_k+i\gamma)}{\sinh(\lambda_k-i\gamma)}\right]^L = -e^{-2i\varphi} \prod_{l=1}^r\frac{\sinh\half[\lambda_k-\lambda_l+2i\gamma]}{\sinh\half[\lambda_k-\lambda_l-2i\gamma]}\,.
\label{eq:BAEcan}
\end{equation}

For small system sizes one can check numerically that  the vacuum of the Hamiltonian~\eref{eq:ham} is antiferromagnetic and the corresponding Bethe roots lie on the lines $\Im\, \lambda = \pm \pi/2$. 
In the following we shall consider low energy solutions of the form
\begin{equation*}
\{\lambda_j\}_{j=1,\dots,r} = \{ \lambda_{0j} -i\pi/2 \}_{j=1, \dots, r_0}\cup
\{ \lambda_{1j} + i\pi/2 \}_{j=1, \dots,r_1}\,,
\end{equation*}
with $\lambda_{aj}$ real.
The logarithmic form of the BAE for this type of solutions is
\begin{equation} \label{eq:BAE2}
  \fl\qquad Lp(\lambda_{aj}) = 2\pi I_{aj} \corr{-} 2\vphi
  - \sum_{b=0,1}\sum_{\ell=1}^{r_b} \theta_{a-b}(\lambda_{aj}-\lambda_{b\ell}) \,,
\end{equation}
where $I_{aj}$ are the Bethe integers with $I_{aj} \in (r_a-1)/2 + \mathbb{Z}$.
The momentum and scattering phases are
\begin{equation} \label{eq:def-p-theta}
  \fl\qquad p(\lambda) = \phi_{\pi/2-\gamma}(\lambda) \,,
  \qquad
  \theta_0(\lambda) = \phi_\gamma(\lambda/2) \,,
  \qquad
  \theta_{\pm 1}(\lambda) = -\phi_{\pi/2-\gamma}(\lambda/2)
\end{equation}
and we have defined
\begin{equation} \label{eq:defphi}
  \phi_\alpha(\lambda) := 2 {\rm Arctan} \left(
  \tanh \lambda \ {\rm cotan}\ \alpha
  \right)
  = -i \log \frac{\sinh(i\alpha-\lambda)}{\sinh(i\alpha+\lambda)} \,.
\end{equation}
The function $\phi_\alpha$ is analytic on the strip $|\Im\, \lambda|<\alpha$,
and the properties of $\phi_\alpha$ that we shall need in the subsequent calculations
are given in the Appendix.

The total momentum and energy
can be written as:
\begin{equation} \label{eq:PE}
  \fl \qquad P = \corr{2\varphi}+\sum_{a,j} \phi_{\pi/2-\gamma}(\lambda_{aj}) \,,
  \qquad
  E = -\sin 2\gamma \ \sum_{a,j} \phi'_{\pi/2-\gamma}(\lambda_{aj}) \,.
\end{equation}
The {\it quasi-momentum} associated to the quasi-shift $\wt\tau$ is
defined as:
\begin{equation} \label{eq:S}
  \fl \qquad K :=  \log \frac{\Lambda(\pi/2)}{\Lambda(0)}
  = \sum_{a,j} (-1)^a k(\lambda_{aj})\,,\qquad
	k(\lambda):= \log \frac{\cosh\lambda + \sin\gamma}{\cosh\lambda - \sin\gamma} \,.
\end{equation}
Note that, under the exchange of $\{\lambda_{0j}\}$ and $\{\lambda_{1j}\}$,
$P$ and $E$ are even, whereas $K$ is odd.
Therefore, we shall restrict in the following for definiteness  to configurations of Bethe roots with $r_0\leq r_1$.

\subsection{Bethe integers}

Our aim in this paper is to study the solutions of the BAE corresponding to a vacuum with holes.
 We expect (at least for certain values of $\gamma$) precisely these solutions to carry the energy and momentum quanta, because
the complex  solutions usually carry the spin quanta. The comprehensive description of low energy complex solutions is left for future work.
The ground state solution is fixed by the following configuration of Bethe integers:
\begin{equation}
  I_{a1}, \dots, I_{ar_a} =
  -\frac{r_a-1}{2}, -\frac{r_a-3}{2}, \dots, \frac{r_a-1}{2} \,,
\end{equation}
where $r_0=r_1=L/2$. In analogy to the XXZ case~\cite{Alcaraz:1987zr},
we restrict to hole configurations obtained by two procedures.

First, one can remove (add) some roots from (to) the ground state:
\begin{equation}
  r_a = L/2 - m_a \,.
\end{equation}
Since the total magnetisation is $S^z=m_0+m_1$, these are called magnetic
excitations. We introduce the even and odd ``magnetic charges''
\begin{equation} \label{eq:def-m}
  m:= m_0+m_1\geq 0 \,,
  \qquad
  \wt m := m_0-m_1 \geq 0\,,
\end{equation}
which we require to be non-negative.
It is important to realize that $\wt m$ is not the eigenvalue of a conserved charge.
Rather, we shall derive the relation $\wt m \propto K \log L$ which holds only
in the continuum limit.

Secondly, one can  shift all the $I_{aj}$ by an integer
$e \in \mathbb{Z}$: these are called electric excitations. A combined electro-magnetic
excitation corresponds to the configuration:
\begin{equation}
  I_{a1}, \dots, I_{ar_a} =
  -\frac{r_a-1}{2}+e, -\frac{r_a-3}{2}+e, \dots, \frac{r_a-1}{2}+e \,.
\end{equation}
Thus, the Bethe integer configuration is determined by three integer numbers $(m,\wt m, e)$,
with $m \equiv \wt m \mod 2$.


\section{Finite size NLIE with singular kernels}\label{sec:NLIE1}



\subsection{Counting functions}
\label{sec:cf}

The BAE~\eref{eq:BAE2} and the conserved quantities~(\ref{eq:PE}\,--\,\ref{eq:S})
can be re-expressed in terms of the counting functions. For
$|\Im\, \lambda|<\min(2\gamma,\pi/2-\gamma)$,
we define:
\begin{equation} \label{eq:def-Z}
  Z_a(\lambda):= L p(\lambda) \corr{+} 2\vphi + \sum_{b,\ell}
  \theta_{a-b}(\lambda-\lambda_{b\ell}) \,,
\end{equation}
so that the BAE~\eref{eq:BAE2} simply read 
\begin{equation}
  Z_a(\lambda_{aj}) = 2\pi I_{aj} \,.
\end{equation}
The limiting values of $Z_a$  are given by~\eref{eq:phiinf}:
\begin{equation} \label{eq:Zinf}
  Z_a(\pm\infty) = \pm \left\{
  \frac{\pi}{2}[L-(-1)^a \wt{m}]
  + \left(2\gamma-\frac{\pi}{2}\right) m
  \right\} \corr{+} 2\vphi \,.
\end{equation}
The function $[1+(-1)^{r_a}\exp(iZ_a)]$ vanishes at $\lambda_{aj}$,
but it can have additional real roots: these are called ``holes'',
and denoted $\eta_{aj}$. The corresponding Bethe integers
are denoted $I_{h,aj}$, and we have BAE for holes:
\begin{equation}
  Z_a(\eta_{aj}) = 2\pi I_{h,aj} \,,
  \qquad \text{with} \quad I_{h,aj} \in \frac{r_a-1}{2} + \mathbb{Z} \,.
\end{equation}

\begin{myfig}
  \includegraphics[scale=1]{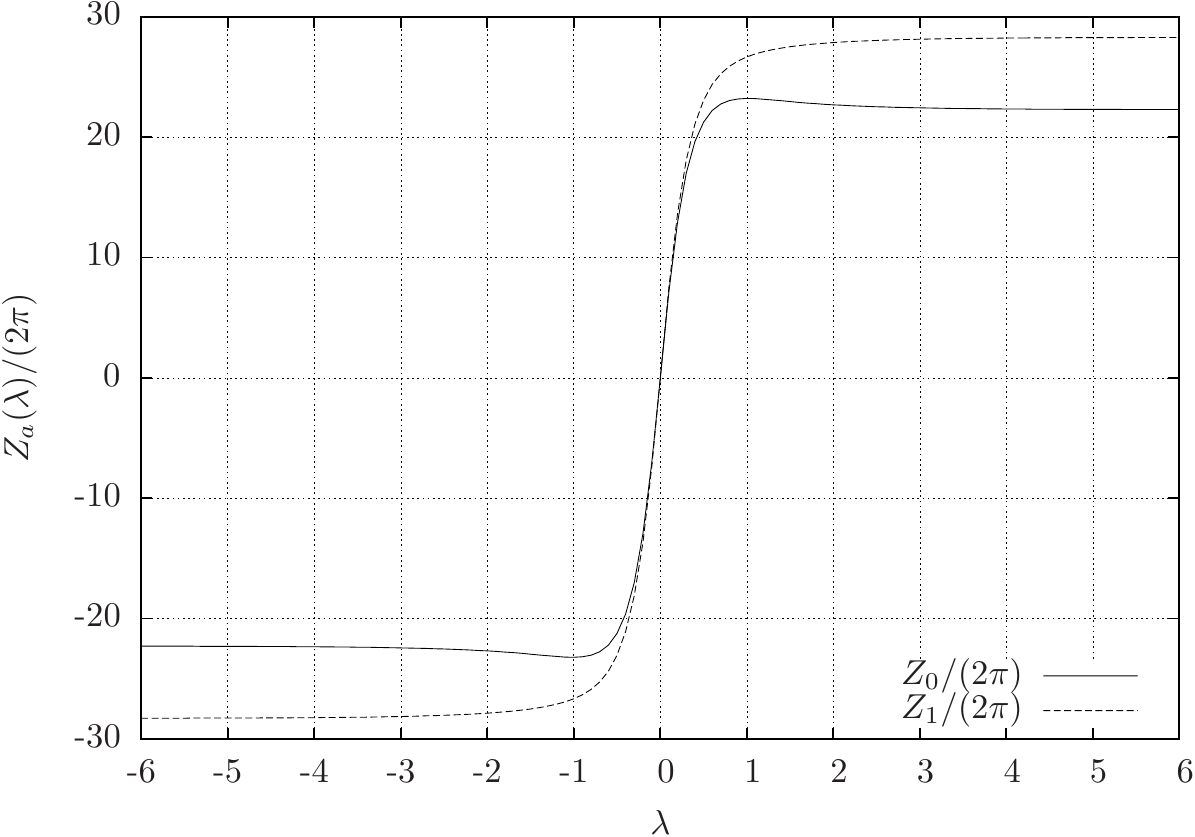}
  \caption{The counting functions $Z_0$ and $Z_1$ at finite size $L=100$, for
    $\gamma=1.24$, $\vphi=0$, $e=0$, $m=2$, $\wt{m}=12$.}
  \label{fig:Z}
\end{myfig}

\begin{myfig}
  \includegraphics[scale=1]{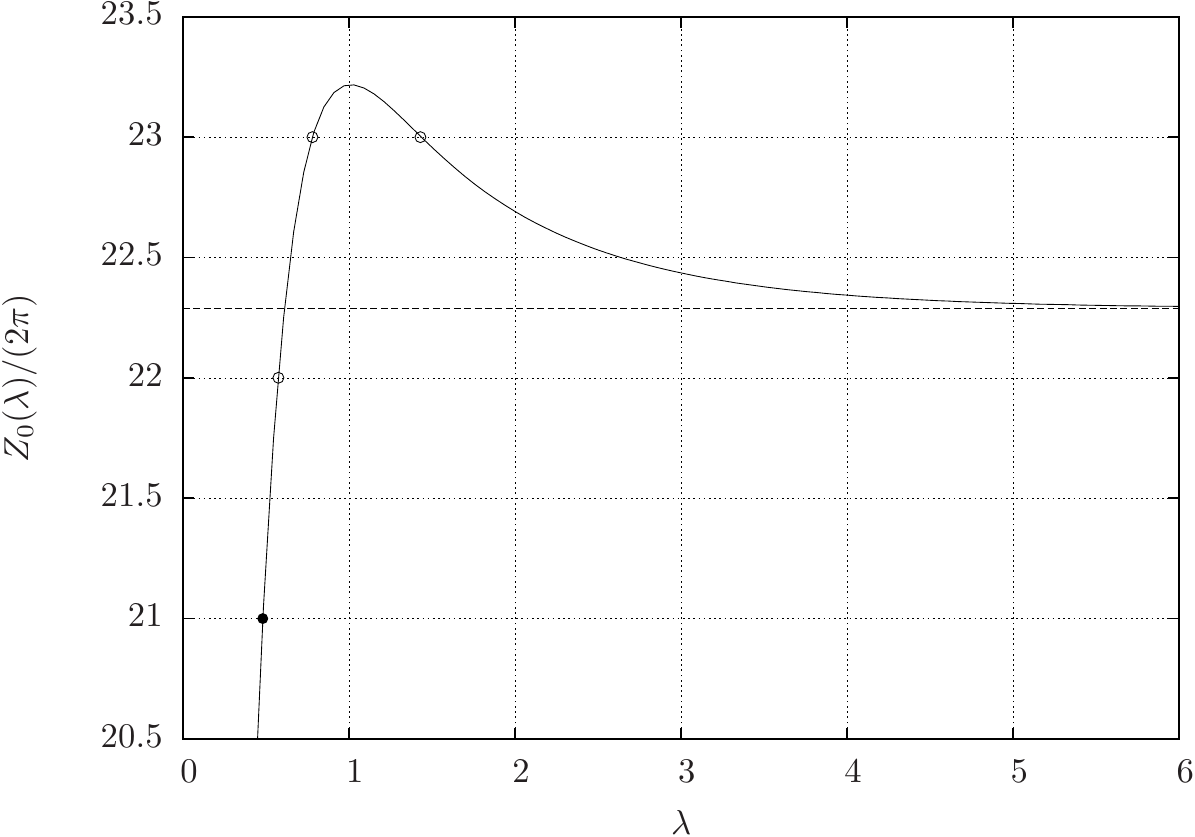}
  \caption{The vicinity of the maximum of $Z_0$ at finite size $L=100$, for
    $\gamma=1.24$, $\vphi=0$, $e=0$, $m=2$, $\wt{m}=12$. Full dots represent the
    Bethe roots $\lambda_{0j}$,
    and empty dots represent the holes $\eta_{0j}$. The dotted line
    shows the value $Z_0(\infty)$.}
  \label{fig:Z-zoom}
\end{myfig}

Let us now do a numerical experiment: we take $L=100$, fix the integers $(m,\wt m,e)$, and solve
the BAE~\eref{eq:BAE2}.
Using the numerical values of the $\lambda_{aj}$, we can compute
the functions $Z_a(\lambda)$~\eref{eq:def-Z}:
see Figures~\ref{fig:Z}\,--\,\ref{fig:Z-zoom}. We then observe the
following facts.
\begin{enumerate}
\item The function $Z_0$ has two extrema, whereas $Z_1$ is increasing. 
\item
  For each Bethe integer in the range $2\pi \max (\pm I_{aj})< 2\pi I_{h,aj}< \pm Z_a(\pm\infty)$,
  we have an {\it ordinary} hole, in the region where $Z'_a>0$.
\item For each Bethe integer in the range $\pm Z_0(\pm \infty)<2\pi I_{h,0j}< \max (\pm Z_0)$,
  we have a pair of {\it extraordinary} holes, with both signs of $Z'_0$.
\end{enumerate}
Thus, the number of positive/negative ordinary holes for $Z_a$ are, respectively,
\begin{equation} \label{eq:Nh}
  N^{\pm}_{h,a} = \left\lfloor \frac{\pm Z_a(\pm\infty)}{2\pi} - \max( \pm I_{aj}) \right\rfloor
  = \left\lfloor \frac{\gamma m\pm\vphi}{\pi} \mp e + \half \right\rfloor := N^\pm_h \,,
\end{equation}
and the number of pairs of positive/negative extraordinary holes for $Z_0$ is, respectively
\begin{equation*}
\wt{N}_h^\pm = \left\lfloor \frac{\max (\pm Z_0)}{2\pi} - \max (\pm I_{aj}) \right\rfloor \,.
\end{equation*}
We denote by $N_h$ (resp. $\wt N_h$)
the total number of ordinary (resp. extraordinary) holes.


\subsection{Non-Linear Integral Equations}
\label{sec:NLIE-finite}


Following~\cite{Destri:1997yz}, we shall reformulate the BAE~\eref{eq:BAE2} as
non-linear integral equations (NLIE) for the counting functions $Z_a$.
This reformulation is most suited for taking the scaling limit.
The non-linear part of the equations involves the functions
\begin{equation}
  \fl\qquad U_a(\lambda) := \log \left[
    1 + (-1)^{r_a} e^{iZ_a(\lambda)}
    \right] \,,
  \qquad 
  \Ub_a(\lambda) := \log \left[
    1 + (-1)^{r_a} e^{-iZ_a(\lambda)}
    \right] \,,
\end{equation}
where we have used the principal determination of the logarithm, the cut-line
being the negative axis.
The functions $U_a(\lambda)$ and $\Ub_a(\lambda)$ will stay away from the branch cut in the domains where $\Im[Z_a(\lambda)]>0$ and, respectively
 $\Im[Z_a(\lambda)]<0$, since the arguments of the logarithms have positive real part.
Hence, their imaginary parts are restricted to the domain
\begin{equation}
-\pi/2< \Im\, U_a(\lambda)\,, \Im\, \Ub_a(\lambda)<\pi/2\,.
\label{eq:restr}
\end{equation}

We can define some integration paths $\Gamma_a$ in the complex plane, on which the $U_a$
are well-defined. First, since $Z_1'(\lambda)>0$ for real $\lambda$, we can take
$\Gamma_1= i\delta + \mathbb{R}$, with $0<\delta<\min(\gamma,\pi-2\gamma)$ finite,
but small enough so that
$\Im[Z_1(\lambda)]>0$ on $\Gamma_1$. For $Z_0$, the path has to be
in the upper half-plane in the vicinity of $Z_0'>0$, and in the lower
half-plane for $Z_0'<0$. The two paths $\Gamma_0$ and $\Gamma_1$ are depicted
in~\figref{contour}.
Similarly, $\Ub_a$ is well-defined on the contour $\Gammab_a$  conjugate to $\Gamma_a$.

\begin{myfig}
  \includegraphics[scale=0.8]{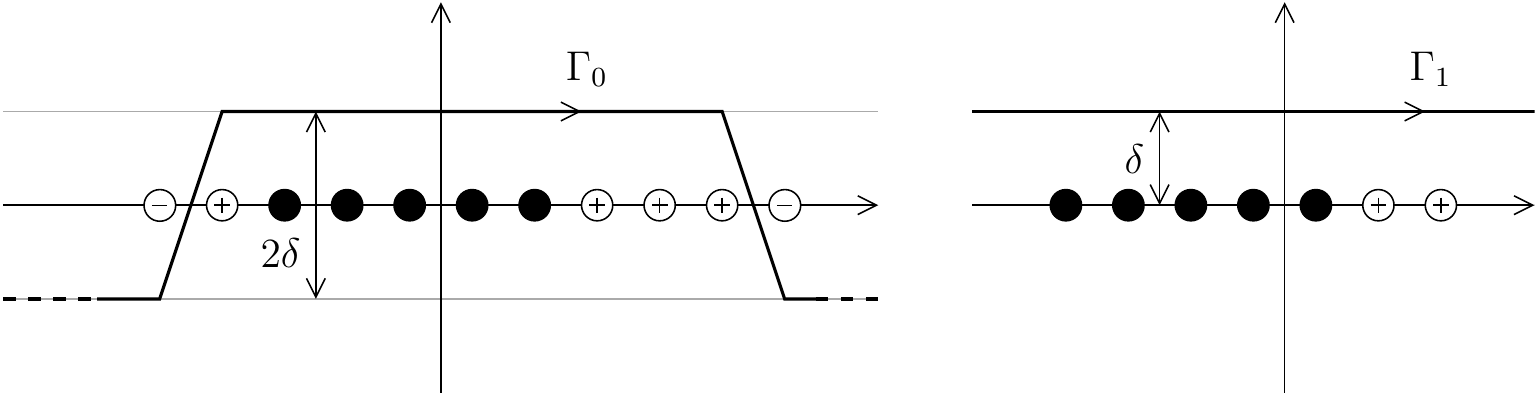}
  \caption{
    Integration paths used in the summation identity~\eref{eq:sum},
    for $\wt m>0$.
    The Bethe roots $\lambda_{aj}$ are depicted by black dots,
    and the holes $\eta_{aj}$ are depicted by white dots, and
    the sign $\nu_{aj}$ is indicated.
    }
  \label{fig:contour}
\end{myfig}

We can now state the basic identity expressing sums over the Bethe roots
in terms of the $\eta_{aj}$ and $U_a$ (see the Appendix for a proof):
\begin{eqnarray}
  \fl\qquad \sum_j f(\lambda_{aj}) &=&
  \int \frac{d\mu}{2\pi} f(\mu) Z'_a(\mu)
  -\sum_j \nu_{aj} f(\eta_{aj}) \nn\\
  \fl&& - \frac{1}{2i\pi} \left[
    \int_{\Gamma_a} d\mu \ f(\mu) U'_a(\mu)
    - \int_{\overline\Gamma_a} d\mu \ f(\mu) \overline U'_a(\mu)
    \right] \,,
   \label{eq:sum}
\end{eqnarray}
which holds if $f$ is a smooth function for real $\lambda$ that increases slow enough at infinity.
The numbers $\nu_{aj}$ are signs, defined as
\begin{equation}
  \nu_{aj} := {\rm sign} \, Z'_a(\eta_{aj}) \,.
\end{equation}
They appear in~\eref{eq:sum} because the integration contour $C_a=(-\Gamma_a) \cup \Gammab_a$
encloses some holes in the clockwise direction, and other holes in the
anti-clockwise direction.

Applying~\eref{eq:sum} to $Z'_a$, we get:
\begin{eqnarray}
  \fl\qquad&&
  \sum_b \left[\left( \delta_{ab} \myid - \frac{\kker_{a-b}}{2\pi} \right) \star Z'_b\right](\lambda)
  = L p'(\lambda) -\sum_{b,\ell} \nu_{b\ell} \kker_{a-b}(\lambda-\eta_{b\ell}) \nn \\
  \fl&&\qquad\qquad - \frac{1}{2i\pi} \sum_b \left[
    \int_{\Gamma_b} d\mu \ \kker_{a-b}(\lambda-\mu) U'_b(\mu)
    - \int_{\Gammab_b} d\mu \ \kker_{a-b}(\lambda-\mu) \Ub'_b(\mu)
    \right] \,,
  \label{eq:NLIE1}
\end{eqnarray}
where $\star$ denotes the convolution product (see~\eref{eq:convol}),
and we have introduced the kernels
\begin{equation}
  \kker_{a-b} := \theta'_{a-b} \,.
\end{equation}
From~(\ref{eq:def-p-theta}, \ref{eq:FT}) we compute the Fourier transforms
\begin{equation}
  \wh{\kker}_0(\omega) = \frac{2\pi \sinh(\pi-2\gamma)\omega}{\sinh \pi\omega} \,,
  \qquad
  \wh{\kker}_{\pm 1}(\omega) =  -\frac{2\pi \sinh 2\gamma\omega}{\sinh \pi\omega} \,.
\end{equation}
Let us first deal with the even part of eq.~\eref{eq:NLIE1}. 
The even kernel and its inverse read:
\begin{eqnarray}
  \wh{\kker}_\even(\omega) &:=& (\wh{\kker}_0 + \wh{\kker}_1)(\omega)
  = \frac{2\pi \sinh(\pi/2-2\gamma)\omega}{\sinh \pi\omega/2} \,,
  \label{eq:Kev} \\
  1+ \wh J_\even(\omega) &:=& \frac{2\pi}{2\pi-\wh {\kker}_\even(\omega)}
  =  \frac{\sinh \pi\omega/2}{2 \sinh \gamma\omega \ \cosh(\pi/2-\gamma)\omega} \,.
  \label{eq:Jev} 
\end{eqnarray}
Summing~\eref{eq:NLIE} over $a$,
convolving with $(1+J_\even)$, and then integrating with respect to $\lambda$,
we get
\begin{eqnarray}
  \fl \quad Z_\even(\lambda) &=& 2 L \sigma(\lambda) + 2C
  - \sum_{b,j} \nu_{bj} H_\even(\lambda-\eta_{bj}) \nn \\
  \fl && - \frac{1}{2i\pi} \sum_b \left[
    \int_{\Gamma_b} d\mu \ H_\even(\lambda-\mu) U'_b(\mu)
    - \int_{\overline\Gamma_b} d\mu \ H_\even(\lambda-\mu) \overline U'_b(\mu)
    \right] \,,
  \label{eq:NLIE-even}
\end{eqnarray}
where $Z_\even := Z_0 + Z_1$, the bulk source term is
\begin{equation}
  \sigma(\lambda) := [(\myid + J_\even) \star p](\lambda)
  = 2 {\rm Arctan} \, \left[ \tanh \frac{\pi\lambda}{2(\pi-2\gamma)} \right] \,,
\end{equation}
and the integrated kernel is the odd function
\begin{equation} \label{eq:Hev}
  H_\even(\lambda) := 2\pi \int_0^\lambda d\mu \ J_\even(\mu) \,.
\end{equation}
The integration constant $C$ in~\eref{eq:NLIE-even} can be easily computed from
the boundary conditions at $\lambda \to \pm\infty$.
From the definition of $N_h^+$, we have
$Z_a(+\infty)-2\pi(I_{a,{\rm max}}+N_h^+) \in [0,2\pi[$,
which gives $Z_a(+\infty)-\pi r_a-2\pi(e+N_h^+) \in [-\pi,\pi[$.
Using a similar argument for $Z_a(-\infty)$, we get
\begin{eqnarray}
  \Im \, U_a(\pm\infty) &=&  \pm (\gamma m-\pi N_h^\pm)+\vphi-\pi e\,,
  \label{eq:Uinf}
\end{eqnarray}
which, as it should, lies in the interval~\eref{eq:restr}.
Plugging these values in~\eref{eq:NLIE-even}, we obtain
\begin{equation}
  C = 2\vphi\,.
\end{equation}

We now turn to the odd part of the NLIE~\eref{eq:NLIE1}.
This involves the odd kernels:
\begin{eqnarray}
  \wh {\kker}_\odd(\omega) &:=& (\wh {\kker}_0 - \wh {\kker}_1)(\omega)
  = \frac{2\pi \cosh(\pi/2-2\gamma)\omega}{\cosh \pi\omega/2} \,,
  \label{eq:Kodd} \\
  1+ \wh J_\odd(\omega) &:=& \frac{2\pi}{2\pi-\wh {\kker}_\odd(\omega)}
  = \frac{\cosh \pi\omega/2}{2 \sinh \gamma\omega \ \sinh(\pi/2-\gamma)\omega} \,.
  \label{eq:Jodd}
\end{eqnarray}
Its treatment will be
very different, due to the singularity of $\wh J_\odd$ at $\omega=0$.
Fourier transforming, we can write:
\begin{eqnarray}
  \fl\quad \wh Z'_\odd(\omega) = \wh J_\odd(\omega) \times \Bigg\{
  && -2\pi \sum_{b,j} (-1)^b \nu_{bj} e^{i\omega\eta_{bj}} \nn \\
  \fl &&- \frac{1}{i} \sum_b (-1)^b \left[
    \int_{\Gamma_b} d\mu\ e^{i\omega\mu} U_b'(\mu)
    - \int_{\overline \Gamma_b} d\mu\ e^{i\omega\mu} \overline U_b'(\mu)
    \right]
  \Bigg\} \,,
  \label{eq:NLIE-odd}
\end{eqnarray}
where $Z_\odd:= Z_0-Z_1$. In the limit $\omega \to 0$, we have
\begin{equation}
  \fl\qquad \wh Z_\odd'(0) = Z_\odd(+\infty) - Z_\odd(-\infty) = -2\pi \wt{m} \,,
  \qquad
  \wh J_\odd(\omega) \mathop{\sim}_{\omega \to 0}
  \frac{\omega^{-2}}{\gamma(\pi-2\gamma)} \,.
\end{equation}
Taking this limit in~\eref{eq:NLIE-odd} yields the consistency conditions
for $n=0,1,2$:
\begin{equation}
  \fl\quad \sum_{b,\ell} (-1)^b \nu_{b\ell} \ \eta_{b\ell}^n
    + \frac{1}{\pi} \sum_b (-1)^b 
    \int_{\Gamma_b} d\mu\, \Im[\mu^n \, U_b'(\mu)] = 
    \left\{ \begin{array}{ll}
      0 & n=0,1 \\
      -\pi\wt{m}/\alpha \ & n=2 \,,
    \end{array} \right.
    \label{eq:cond}
\end{equation}
where we have defined
\begin{equation}
\alpha :=  \frac{\pi}{2\gamma(\pi-2\gamma)}\,.
\label{eq:alpha}
\end{equation}
These can be gathered into a single equation for any $\lambda$:
\begin{eqnarray} \label{eq:cond-poly}
  \fl \quad && \sum_{b,j} (-1)^b \nu_{bj} (\lambda-\eta_{bj})^2  \\
  \fl \quad && + \frac{1}{2i\pi} \sum_b (-1)^b \left[
    \int_{\Gamma_b} d\mu\ (\lambda-\mu)^2\ U_b'(\mu)
    - \int_{\Gammab_b} d\mu\ (\lambda-\mu)^2\ \Ub_b'(\mu)
    \right] = -\frac{\pi\wt{m}}{\alpha} \,. \nn
\end{eqnarray}
The Fourier transform of the singular kernel $\wh J_\odd$ can be defined in various ways.
If we use the principal value prescription:\footnote{The principal value can be computed by averaging the integrals over $\mathbb{R}\pm i\epsilon$.}
\begin{eqnarray}
  J_\odd(\lambda) &:=& \fint \frac{d\omega}{2\pi} e^{-i\lambda\omega} \wh J_\odd(\omega)\,, \qquad
	H_\odd(\lambda) := 2\pi \int_0^\lambda d\mu \ J_\odd(\mu) \,,
\end{eqnarray}
then $J_\odd$ is even and $H_\odd$ is odd. Their asymptotic for large $\lambda$, up to exponentially small terms, is determined by the pole at $\omega=0$
\begin{eqnarray} 
  && J_\odd(\lambda) \sim
  \mp \alpha\lambda/ \pi \,, \qquad
   H_\odd(\lambda) \sim
  \mp(\alpha \lambda^2 - \beta) \,,\qquad \lambda\to\pm\infty\,,
   \label{eq:lim-JHodd}
\end{eqnarray}
where, although irrelevant in the following, $\beta=(\pi^2 \alpha-5\pi)/6$.
After inverse Fourier transforming and integrating eq.~\eref{eq:NLIE-odd}, we get
\begin{eqnarray}\label{eq:zodd_fin}
  \fl \quad Z_\odd(\lambda) &=& 2\wt C -\sum_{b,j} (-1)^b \nu_{bj} H_\odd(\lambda-\eta_{bj}) \\ \nn
  \fl && - \frac{1}{2i\pi} \sum_b (-1)^b \left[
    \int_{\Gamma_b} d\mu\ H_\odd(\lambda-\mu) U'_b(\mu)
    - \int_{\overline\Gamma_b} d\mu\ H_\odd(\lambda-\mu) \overline U'_b(\mu)
    \right] \,,
\end{eqnarray}
where $\wt C$ is an integration constant. The consistency condition~\eref{eq:cond-poly}
actually ensures that the solution $Z_\odd(\lambda)$ has finite limits at $\pm \infty$.
Taking $\lambda \to \pm \infty$ and using~\eref{eq:cond-poly}, we find that
\begin{equation*}
  \wt C=0 \,.
\end{equation*}

Finally, we can recombine~\eref{eq:NLIE-even} and~\eref{eq:NLIE-odd}. Introducing
\begin{eqnarray*}
  J_{a-b} &:=& \half [J_\even + (-1)^{a-b} J_\odd] \,, \qquad
  H_{a-b} := \half [H_\even + (-1)^{a-b} H_\odd] \,,
\end{eqnarray*}
and recalling the BAE for holes, we have the following system of equations:
\begin{equation}
  \fl \boxed{ \begin{array}{rcl}
      Z_a(\lambda) &=& \displaystyle L \sigma(\lambda) + 2\vphi
      -\sum_{b,j} \nu_{bj} H_{a-b}(\lambda-\eta_{bj})  \\
      && \displaystyle - \frac{1}{2i\pi} \sum_b \left[
        \int_{\Gamma_b} d\mu\ H_{a-b}(\lambda-\mu) U'_b(\mu)
        - \int_{\overline\Gamma_b} d\mu\ H_{a-b}(\lambda-\mu) \overline U'_b(\mu)
        \right]  \\
      Z_a(\eta_{aj}) &=& 2\pi I_{h,aj} \,.
  \end{array}}
  \label{eq:NLIE}
\end{equation}
The NLIE~\eref{eq:NLIE} are exact for finite $L$ and they hold for $|\Im\,\lambda|<\min (2\gamma,\pi/2-\gamma)$. It is important to realize that, at this stage,
integration by parts in the first equation of~\eref{eq:NLIE} is not possible,
because the kernels $H_{a-b}(\lambda)$ diverge at $\pm \infty$: see~\eref{eq:lim-JHodd}.


\subsection{Equations for conserved charges}


In this section, shall
express conserved quantities which are defined as sums over roots
\begin{equation*}
  V_\even := \sum_{a,j} v(\lambda_{aj}) \,,
  \qquad
  W_\odd := \sum_{a,j} (-1)^a \ w(\lambda_{aj}) \,,
\end{equation*}
where $v(\lambda)$ and $w(\lambda)$ are smooth functions, in terms of $Z_a$ and sums over holes.
Using eq.~\eref{eq:sum} and the NLIE~\eref{eq:NLIE}, we get
\begin{eqnarray}
  \fl \qquad V_\even &=& 2 L v_\infty
  + \sum_{b,\ell} \nu_{b\ell} \ v_h(\eta_{b\ell}) \nn \\
  \fl && + \frac{1}{2i\pi} \sum_b \left[
    \int_{\Gamma_b} d\lambda\ v_h(\lambda) U'_b(\lambda)
    - \int_{\Gammab_b} d\lambda\ v_h(\lambda) \Ub'_b(\lambda)
    \right] \,, \label{eq:Wev}
\end{eqnarray}
where 
\begin{eqnarray*}
  v_\infty &:=& \int \frac{d\lambda}{2\pi} \ \sigma'(\lambda) v(\lambda) \,,\qquad
  v_h := -(1+J_\even) \star v \,.
\end{eqnarray*}
The first term in~\eref{eq:Wev} is the bulk value, and the next
terms give the finite-size corrections to $V_\even$.
Since the energy and momentum of a Bethe root are given by
\begin{equation}
  p(\lambda) = \phi_{\pi/2-\gamma}(\lambda) \,,
  \qquad
  \epsilon(\lambda) = -\sin 2\gamma \ \phi'_{\pi/2-\gamma}(\lambda) \,,
\end{equation}
the total energy and momentum read
\begin{eqnarray}
  \fl\qquad P &=& 2\varphi+\sum_{b,\ell} \nu_{b\ell} \ p_h(\eta_{b\ell})
  + \frac{1}{\pi} \sum_b
    \int_{\Gamma_b} d\lambda \ \Im[p_h(\lambda) U'_b(\lambda)] \,, \label{eq:P2}
  \\
  \fl\qquad E &=& 2 L e_\infty
  +  \sum_{b,\ell} \nu_{b\ell} \ \epsilon_h(\eta_{b\ell})
  + \frac{1}{\pi} \sum_b
    \int_{\Gamma_b} d\lambda \ \Im[\epsilon_h(\lambda) U'_b(\lambda)] \,, \label{eq:E2}
\end{eqnarray}
where the energy and momentum of a hole (in the density approximation) are
\begin{equation}\label{eq:phlatt}
  p_h(\lambda) = -\sigma(\lambda)
\,,
  \qquad
  \epsilon_h(\lambda) = v_F/\cosh \frac{\pi\lambda}{\pi-2\gamma} \,,
\end{equation}
and we have introduced the ``Fermi velocity'' $v_F:= \pi\sin2\gamma/(\pi-2\gamma)$.
Note that we have the exact relation
  $\epsilon_h = v_F \cos p_h$
and hence, at $\lambda \to \pm \infty$, since $p_h \to \mp \pi/2$, the holes have a linear
dispersion relation.

For the odd conserved quantities, we can derive an expression similar to~\eref{eq:Wev}
\begin{eqnarray}
  \fl\qquad W_\odd &=& \sum_{b,\ell} (-1)^b \nu_{b\ell} \ w_h(\eta_{b\ell}) \nn \\
  \fl&& + \frac{1}{2i\pi} \sum_b (-1)^b \left[
    \int_{\Gamma_b} d\lambda\ w_h(\lambda) U'_b(\lambda)
    - \int_{\Gammab_b} d\lambda\ w_h(\lambda) \Ub'_b(\lambda)
    \right] \,,
  \label{eq:Wodd}
\end{eqnarray}
where $w_h := -(1+J_\odd) \star w$.
Hence, we can write the
quasi-momentum \eref{eq:S} as
\begin{equation} \label{eq:S2}
  \fl\qquad K = \sum_{b,\ell} (-1)^b \nu_{b\ell} \ k_h(\eta_{b\ell})
   + \frac{1}{\pi} \sum_b (-1)^b
    \int_{\Gamma_b} d\lambda\ \Im[k_h(\lambda) U'_b(\lambda)] \,,
\end{equation}
where the quasi-momentum of a hole, computed in the Appendix, reads
\begin{equation}
  k_h(\lambda) = \log \left(
  2 \cosh \frac{\pi\lambda}{\pi-2\gamma}
  \right) \,.
	\label{eq:sh}
\end{equation}


\subsection{Higher-order conserved charges}\label{sec:hs}

From~\eref{eq:Lambda}, if we set $u_0:=-\pi/4+\gamma/2$, we can write
\begin{eqnarray}
  \fl \quad \Lambda(u) &=&
  e^{i\vphi} \prod_{a,j} e^{-i \phi_{\gamma/2}[\half(\lambda_{aj}+i\pi a-i\pi/2+2iu)]}
  \times[1+(-1)^{r_0}e^{-iZ_0(2iu_0-2iu)}] \,, \\
  \fl \quad \Lambda(u+\pi/2) &=&
  e^{i\vphi} \prod_{a,j} e^{-i \phi_{\gamma/2}[\half(\lambda_{aj}+i\pi a+i\pi/2+2iu)]}
  \times[1+(-1)^{r_1}e^{-iZ_1(2iu_0-2iu)}] \,.
\end{eqnarray}
Taking the logarithm we get
\begin{eqnarray*}
  \fl \quad \log\Lambda(u) &=&
  i\vphi -i\sum_{a,j} \phi_{\gamma/2} \left[\half(\lambda_{aj}+i\pi a-i\pi/2+2iu) \right]
  +\Ub_0(2iu_0-2iu) \,, \\
  \fl \quad \log\Lambda(u+\pi/2) &=&
  i\vphi - i\sum_{a,j} \phi_{\gamma/2} \left[\half(\lambda_{aj}+i\pi a+i\pi/2+2iu) \right]
  +\Ub_1(2iu_0-2iu) \,.
\end{eqnarray*}
For the even (free energy) and odd combinations:
\begin{equation}
  \fl \quad F(u):= -\log[\Lambda(u)\Lambda(u+\pi/2)] \,,
  \qquad
  G(u):= \log[\Lambda(u+\pi/2)/\Lambda(u)] \,,
\end{equation}
one arrives with the help of eq.~\eref{eq:phidbl} at
\begin{eqnarray}\label{eq:fgdef}
  \fl \quad F(u) &=& -2i\vphi -i\sum_{a,j} \phi_{\pi/2-\gamma}(\lambda_{aj}+2iu)
  - (\Ub_0+\Ub_1)(2iu_0-2iu) \,, \\
  \fl \quad G(u) &=& \sum_{a,j} (-1)^a 
  k(\lambda_{aj}+2iu)
  + (\Ub_0-\Ub_1)(2iu_0-2iu) \,. \nn
\end{eqnarray}
If we now restrict to\footnote{The values $0<\gamma<\pi/6$ can be treated in
a similar way by changing variable $u \to \gamma-u$.}
\begin{equation}\label{eq:strip}
  \pi/6<\gamma<\pi/2 \,, \qquad
  \gamma-\pi/2< \Re\,2u<0 \,.
\end{equation}
then the summands in eqs.~\eref{eq:fgdef} are analytic functions and $\gamma -\pi/2< \Im\,2i(u_0-u)< 0$.
Hence, we can use eqs.~\eref{eq:Wev} and \eref{eq:Wodd} to express $F(u)$ and $G(u)$ in terms of the solution of
the NLIE \eref{eq:NLIE} as follows:
\begin{eqnarray}\label{eq:fsol}
  \fl \quad F(u) &=& -2i\vphi +2L f_\infty(u) -i \sum_{b,\ell} \nu_{b\ell}\ p_h(\eta_{b\ell}+2iu)
  - (\Ub_0+\Ub_1)(2iu_0-2iu) \nn \\
  \fl \quad &&- \frac{1}{2\pi} \sum_b \left[
    \int_{\Gamma_b} d\lambda \ p_h(\lambda+2iu) U'_b(\lambda)
    -\int_{\Gammab_b} d\lambda \ p_h(\lambda+2iu) \Ub'_b(\lambda)
    \right] \,, \\
  \fl \quad G(u) &=& \sum_{b,\ell} (-1)^b \nu_{b\ell}\ k_h(\eta_{b\ell}+2iu)
  +(\Ub_0-\Ub_1)(2iu_0-2iu) \nn \\
  \fl \quad &&+ \frac{1}{2i\pi} \sum_b (-1)^b  \left[
    \int_{\Gamma_b} d\lambda \ k_h(\lambda+2iu) U'_b(\lambda)
    -\int_{\Gammab_b} d\lambda \ k_h(\lambda+2iu) \Ub'_b(\lambda)
    \right] \,,
		\label{eq:gsol}
\end{eqnarray}
where $f_\infty(u):=-i (\sigma'\star p)(2iu)$ is the bulk free energy density.
Unless there are complex solutions of the BAE in the strip~\eref{eq:strip}, the functions $F(u)$ and $G(u)$ will be analytic there.

From $F(0)$, $F'(0)$ and $G(0)$, one recovers $P$, $E$ and $K$. Higher
derivatives of $F$ and $G$ give the eigenvalues of
all the (local) conserved charges which commute with $H$.


\section{Scaling limit of the NLIE}\label{sec:NLIE2}


\subsection{Definitions}


\begin{myfig}
  \includegraphics[scale=1]{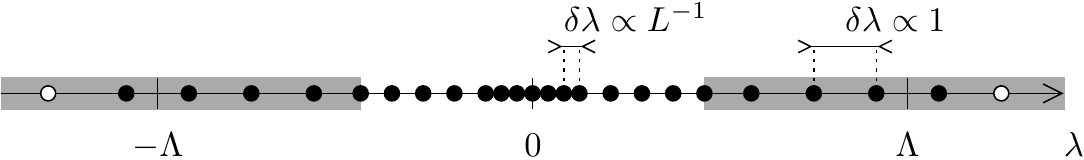}
  \caption{Positions of Bethe roots and holes for a large system size $L$.
  For $|\lambda| \ll \Lambda$, the spacing between Bethe roots is of order $1/L$,
  whereas for $|\lambda| \simeq \Lambda$, it is finite.
  The shaded regions around $\pm \Lambda$ represent the scaling regime
  described by the shifted functions $Z_a^\pm$.}
  \label{fig:scaling}
\end{myfig}

We shall now focus on the excitations concentrated at the boundaries of the Fermi sphere of roots
where the dispersion relation for the holes linearises and the gap closes as $\mathcal{O}(1/L)$
in the $L\to\infty$ limit.
This is the ``conformal regime'' described by a CFT.\footnote{Instead, one could engineer a gap in the dispersion relation of the holes with $|\lambda|<\Theta$ by adding an additional \emph{imaginary} staggering $2i\Theta$ to the spectral parameters in fig.~\ref{fig:stag}, see \cite{Destri:1987ug, Reshetikhin:1993wm}. This is the ``massive regime'' described by a massive integrable quantum field theory.}
The latter was identified with the $\mathrm{SL}(2,\mathbb{R})/\mathrm{U}(1)$ black hole sigma model; $\mathcal{O}(1/L)$ corrections to the energy~\eref{eq:E2} then
give the central charge and  spectrum of this CFT \cite{Ikhlef:2011ay}.
Before we are able to reproduce these corrections
we  need to take the scaling limit of the NLIE~\eref{eq:NLIE}.

We define the scaling limit as
\begin{equation} \label{eq:scaling}
  L \to \infty \,,
  \qquad (m,e,K) \ \text{fixed.}
\end{equation}
Numerical inspection shows that $K \propto \wt m / \log L$ for large $L$.
Hence, the scaling limit  requires taking $\tilde{m}\to\infty$.
We shall confirm this behaviour in Sec.~\ref{sec:source-terms}.

Like in the XXZ case~\cite{Klummpe:1991vs, Destri:1992qk}, from the asymptotic behaviour
\begin{equation}\label{eq:asymptsig}
  \sigma(\lambda) \mathop{\sim}_{\lambda \to \pm \infty}
  \pm \left[ \pi/2
    - 2 e(\pm \lambda)
    \right] \,,\qquad
		e(\lambda):=\exp \left(-\frac{\pi\lambda}{\pi-2\gamma}\right)
\end{equation}
we find that the $\lambda$ dependent part of the source term $L\sigma(\lambda)$ of~\eref{eq:NLIE} will be  of order one
if $\lambda$ remains near the ``Fermi levels'' $\pm \Lambda$ with
\begin{equation}
  \Lambda := \frac{\pi-2\gamma}{\pi} \log L \,.
\end{equation}
In the limit~\eref{eq:scaling}, the Bethe roots and holes arrange
as follows (see~\figref{scaling}).
For $|\lambda| \ll \Lambda$, the spacing between roots is of order $1/L$, and the
corresponding contributions to conserved quantities are contained in the
bulk term $2Lv_\infty$ of~\eref{eq:Wev}. Actually, this regime is well described
by the linear approximation and the Wiener-Hopf approach of Yang and Yang~\cite{Yang:1966sa}.
For $|\lambda| \simeq \Lambda$,
the roots are of the form $\lambda_{aj} = \pm(\Lambda + \lambda_{aj}^\pm)$,
where $\lambda_{aj}^\pm$ remains finite as $L$ becomes large,
and similarly for holes. This is the scaling regime which determines the
finite-size corrections (shaded regions in~\figref{scaling}).
Also, as we shall see in the subsequent calculations,
the NLIE for the positive and negatives holes (right and left movers)
decouple in the scaling limit.

Let us then define the shifted counting functions:
\begin{equation} \label{eq:def-Zpm}
  Z_a^\pm(\lambda) := \pm Z_a[\pm(\Lambda+\lambda)] - \pi r_a \,.
\end{equation}
Notice that they have a finite value at $\lambda \to +\infty$
in the limit~\eref{eq:scaling}:
\begin{equation} \label{eq:Zinf-shift}
  Z_a^\pm(+\infty) = 2\gamma m \pm 2\vphi \,.
\end{equation}
Correspondingly, we write
\begin{eqnarray}\nn
  U^+_a(\lambda) &:=&  U_a(\Lambda+\lambda)=\log \left[
  1 + e^{iZ_a^+(\lambda)}
  \right]\,,\\
  U^-_a(\lambda) &:=& \Ub_a(-\Lambda-\lambda)=\log \left[
    1 + e^{iZ_a^-(\lambda)}
    \right]\,,
		\label{eq:scaledU}
\end{eqnarray}
and define $\Ub^\pm_a(\lambda^*):=U^\pm_a(\lambda)^*$, where the operation $(\cdot)^*$  denotes complex conjugation.
The  scaled integration paths $\Omega^\pm_a=\Omega_a$ compatible with  eqs.~\eref{eq:scaledU} are shown in~\figref{contour-shift}.
Their complex conjugate is denoted by $\overline{\Omega}_a$.
From eqs.~\eref{eq:Uinf} we get the boundary conditions
\begin{equation}
\Im\, U^\pm_a(+\infty)=\gamma m-\pi N_h^\pm \pm (\varphi-\pi e)\,,
\label{eq:Uscinf}
\end{equation}
while from the behaviour of the source term $L\sigma(\lambda)$ at $L\to\infty$ and $\lambda$ of order one we expect
$\Im \, Z_a^\pm(i\delta-\infty) = +\infty$, and thus:
\begin{equation}\label{eq:Uscminf}
  U_a^\pm(i\delta -\infty) = 0 \,.
\end{equation}
The shifted Bethe holes and the associated signs are defined as
\begin{equation}
  \{ \eta_{aj} \} = \{ \Lambda + \eta^+_{aj} \} \cup \{ -\Lambda - \eta^-_{aj} \} \,, \qquad
  \nu_{aj}^\pm = {\rm sign} \, {Z^{\pm}_a}'(\eta^\pm_{aj}) \,,
\end{equation}
and the shifted Bethe integers for roots are
\begin{equation}\label{eq:scaledBI}
  I_{aj}^\pm := \pm I_{aj} - r_a/2 \in \pm e  +1/2-\mathbb{N}\,.
\end{equation}
The Bethe integers for holes take values in  the complementary interval in $\mathbb{Z}+1/2$.

\begin{myfig}
  \includegraphics[scale=0.9]{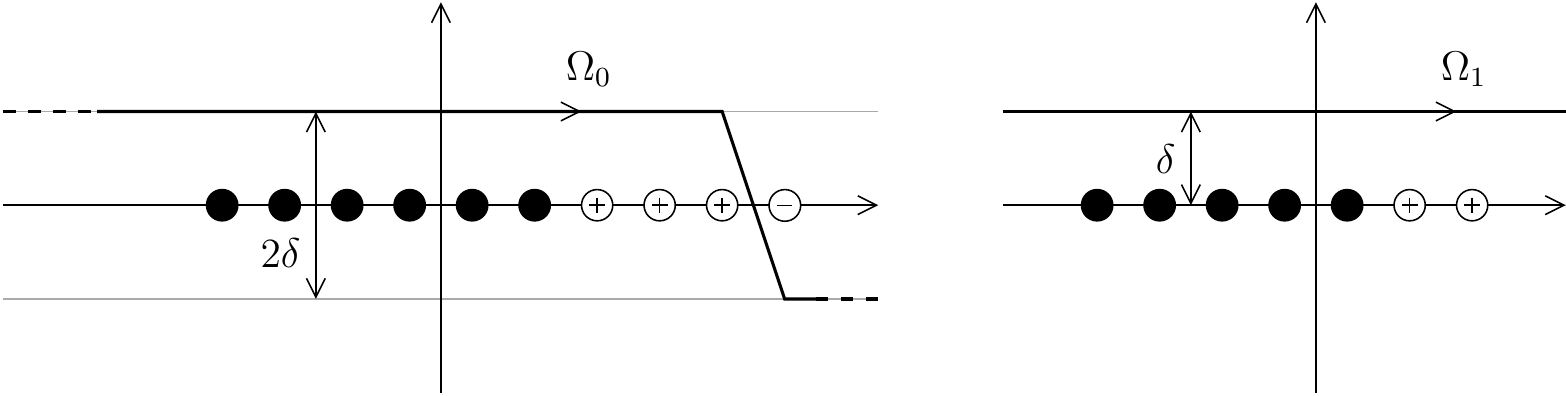}
  \caption{Integration paths for the shifted counting functions $Z_a^\pm$.
    See~\figref{contour} for details.}
  \label{fig:contour-shift}
\end{myfig}


\subsection{Non-Linear Integral Equations}
\label{sec:NLIE-scaling}


For the sake of clarity, we first derive the NLIE for the $Z_a^+$, and
state the analogous results for the $Z_a^-$. Furthermore, like
in~\secref{NLIE-finite}, we separate the discussion of the even and odd (under
the exchange of indices $a=0 \leftrightarrow 1$) parts
of the NLIE.

In the even NLIE~\eref{eq:NLIE-even}, we perform the change of variables
$\lambda \to \Lambda + \lambda$, and then integrate by parts. In the scaling
limit~\eref{eq:scaling}, the negative holes and the negative part of the integral
contribute an additive constant, and we obtain:
\begin{eqnarray}
  \fl\qquad Z_\even^+(\lambda) &=& -4e(\lambda) + 2C^+
  -\sum_{b,\ell} \nu_{b\ell}^+ H_\even(\lambda-\eta_{b\ell}^+) \nn \\
  \fl &&-\frac{1}{i} \sum_b \left[
  \int_{\Omega_b} d\mu\ J_\even(\lambda-\mu) U^+_b(\mu)
  - \int_{\overline{\Omega}_b} d\mu\ J_\even(\lambda-\mu) \Ub^+_b(\mu)
  \right] \,,
  \label{eq:NLIE-ev-scaling}
\end{eqnarray}
where, taking into account eqs.~(\ref{eq:Uscinf}, \ref{eq:Uscminf}), we get
\begin{eqnarray}
  C^+ &=& \frac{\pi}{2} \left(m+\frac{\vphi}{\gamma} \right)
  + \pi\left(1-\frac{\pi}{4\gamma}\right)(N_h^++2e) \,.
\end{eqnarray}

Before we deal with the odd part it is useful to define the modified kernels
\begin{eqnarray}
	\wt{H}_\odd(\lambda) &:=& H_\odd(\lambda) - (\alpha \lambda^2-\beta) \,,\qquad
	\wt{H}'_\odd=2\pi\wt{J}_\odd\,,
\end{eqnarray}
which decay exponentially at $\lambda\to-\infty$.
Now, notice that $Z^+_\odd(\lambda) = Z_\odd (\Lambda+\lambda)+\pi\wt m$ and write
the consistency conditions~(\ref{eq:cond}\,--\,\ref{eq:cond-poly}) as
\begin{eqnarray*} 
  \fl\quad&& \pi \wt m =
  -\sum_{b,\ell} (-1)^b \nu_{b\ell} [\alpha(\lambda-\eta_{b\ell})^2-\beta] \nn \\
  \fl && -\frac{1}{2i\pi} \sum_b (-1)^b \left[
    \int_{\Gamma_b} d\mu \ [\alpha(\lambda-\mu)^2-\beta] U'_b(\mu)
    - \int_{\Gammab_b} d\mu \ [\alpha(\lambda-\mu)^2-\beta] \Ub'_b(\mu)
    \right] \,.
\end{eqnarray*}
Adding this equation to eq.~\eref{eq:zodd_fin}
we get:
\begin{eqnarray}\label{eq:zodd_oneform}
  \fl \qquad && Z_\odd^+(\lambda) = \sum_{b,\ell} (-1)^b \nu_{b\ell}^+
  \ \wt H_\odd(-\lambda+\eta_{bj}^+) \\\nn
  \fl && \qquad + \frac{1}{2i\pi} \sum_b (-1)^b \left[
  \int_{\Omega_b} d\mu\ \wt H_\odd(-\lambda+\mu) (U^+_b)'
  - \int_{\overline{\Omega}_b} d\mu\ \wt H_\odd(-\lambda+\mu) (\Ub^+_b)' \right] \,,
\end{eqnarray}
where, because the kernels $-\wt{H}_\odd(-\lambda)=H_\odd (\lambda)+(\alpha\lambda^2-\beta)$ decay
 exponentially at $\lambda\to+\infty$, the left movers have decoupled.
The message conveyed by this equation is that $Z^+_\odd(\lambda)$ decays exponentially at $\lambda\to+\infty$.
In order to understand the asymptotic behaviour of $Z^+_\odd(\lambda)$ at $\lambda\to-\infty$
we perform the replacement	
\begin{equation*}
  \wt H_\odd(-\lambda) = -\wt H_\odd(\lambda) - 2(\alpha \lambda^2-\beta)
\end{equation*}
in eq.~\eref{eq:zodd_oneform}, which, after integration by parts, brings it to the form
\begin{eqnarray} \label{eq:NLIE-odd-scaling}
  \fl\qquad && Z_\odd^+(\lambda) = 4A_1^+ \lambda - 2A_2^+
  -\sum_{b,\ell} (-1)^b \nu_{b\ell}^+ \ \wt H_\odd(\lambda-\eta_{bj}^+)  \\\nn
  \fl && \qquad  - \frac{1}{i} \sum_b (-1)^b \left[
  \int_{\Omega_b} d\mu\ \wt J_\odd(\lambda-\mu) U^+_b(\mu)
  - \int_{\overline{\Omega}_b} d\mu\ \wt J_\odd(\lambda-\mu) \Ub^+_b(\mu) \right] \,.
\end{eqnarray}
Here  $A^+_n$ are constants defined as 
\begin{equation} \label{eq:def-An}
  \fl \qquad A_n^+ := \alpha \Big\{
  \sum_{b,\ell} (-1)^b \nu_{b\ell}^+ (\eta_{b\ell}^+)^n
  + \frac{1}{\pi} \sum_b (-1)^b
    \int_{\Omega_b} d\mu \ \Im[\mu^n \ (U^+_b)'(\mu)] \Big\} \,.
\end{equation}
The first two terms on the r.h.s.\ of eq.~\eref{eq:NLIE-odd-scaling} give the asymptotic behaviour of $Z^+_\odd(\lambda)$ at $\lambda\to-\infty$.
Notice that the integration by parts was possible because $\wt{H}_\odd(\lambda)U^+_a(\lambda)$ decays (exponentially) at $\lambda\to \pm\infty$.

We can now recombine the NLIE~\eref{eq:NLIE-ev-scaling} and \eref{eq:NLIE-odd-scaling},
and, using the kernels
\begin{eqnarray*}
  \fl \qquad \wt J_{a-b} := \half [J_\even+(-1)^{a-b} \wt J_\odd] \,, \qquad
  \wt H_{a-b} := \half [H_\even+(-1)^{a-b} \wt H_\odd] \,,
\end{eqnarray*}
cast the result in its final form:\footnote{Similar NLIE have appeared previously in the literature \cite{Lukyanov}.}
\begin{equation} \label{eq:NLIE-scaling}
  \fl \boxed{\quad \begin{array}{l}
      Z^+_a(\lambda) = \displaystyle -2e(\lambda) + C^+ + (-1)^a (2A^+_1 \lambda - A^+_2)
      -\sum_{b,\ell} \nu_{b\ell}^+ \wt H_{a-b}(\lambda-\eta_{b\ell}^+) \\
      \qquad \qquad \displaystyle - \frac{1}{i} \sum_b \left[
        \int_{\Omega_b} d\mu\ \wt J_{a-b}(\lambda-\mu) U_b^+(\mu)
        - \int_{\overline{\Omega}_b} d\mu\ \wt J_{a-b}(\lambda-\mu) \Ub_b^+(\mu)
        \right] \,, \\
      \\
     C^+ = \displaystyle \frac{\pi}{2} \left(m+\frac{\vphi}{\gamma} \right)
  + \pi\left(1-\frac{\pi}{4\gamma}\right)(N_h^++2e) \,, \\
      \\
      A_n^+ = \displaystyle \alpha \Big\{
      \sum_{b,\ell} (-1)^b \nu_{b\ell}^+ (\eta_{b\ell}^+)^n
      + \frac{1}{\pi} \sum_b (-1)^b
        \int_{\Omega_b} d\mu
        \ \Im \, \left[\mu^n \ (U^+_b)'(\mu) \right] \Big\}  \,,\\
      \\
      Z_a^+(\eta_{aj}^+) = 2\pi I_{h,aj}^+ \,.
  \end{array}}
\end{equation}

As it is apparent in~\eref{eq:NLIE-scaling}, the most notable effect of the
pole at $\omega=0$ of $\wh J_\odd$, see eq.~\eref{eq:Jodd}, is to create an additional source term proportional
to $\lambda$ in the NLIE. Moreover, for one value of $a$, the source
term $-2e(\lambda)+2|A_1^+|\lambda$
is increasing, whereas for the other value, the source term $-2e(\lambda)-2|A_1^+|\lambda$
has one local maximum: this is reminiscent of the behaviour of $Z_0$ and $Z_1$ observed
numerically for finite system sizes (see~Figs.~\ref{fig:Z}\,--\,\ref{fig:Z-zoom}).

Finally, the minus version of all equations in this section, i.e.\ for the left movers,
can be obtained by replacing all $+$-superscripts with $-$-superscripts and flipping the sign of $e$ and $\varphi$.


\subsection{Constraints and source terms
}
\label{sec:source-terms}

Notice that for $L$ large and $\lambda$ of order one,  the increasing bulk term $L\sigma(\lambda)$ dominates on the r.h.s.\ of $Z_a(\lambda)$ in eq.~\eref{eq:NLIE}.
Hence, in this regime $U_a$ and  $\Ub_a$ get exponentially suppressed with the system size $L$
\begin{equation}
|U_a(\lambda+i\delta)|=|\Ub_a(\lambda-i\delta)|\approx e^{-\delta L \sigma'(\lambda) }\ll 1\,,
\qquad \lambda\in\mathbb{R}\,.
\label{eq:ignored}
\end{equation}
These functions will be of order one precisely in the scaling regime, i.e.\ for $|\lambda|$ of order $\Lambda$. Hence, the following approximation
 is exact up to exponentially small terms in $L$
\begin{eqnarray}\nn
\fl \quad &&\int_{\Gamma_a} d\mu\, f(\mu)U'_a(\mu)-\int_{\bar{\Gamma}_a} d\mu\, f(\mu)\Ub'_a(\mu)\approx  \int_{\Omega_a}d\mu\, \left[f^+(\mu)(U_a^+)'(\mu) +f^-(\mu)(U_a^-)'(\mu)\right]\\
\fl   && \qquad-\int_{\overline{\Omega}_a}d\mu\, \left[f^+(\mu)(\Ub_a^+)'(\mu) +f^-(\mu)(\Ub_a^-)'(\mu)\right]\,,
\label{eq:split}
\end{eqnarray}
where $f^\pm(\mu):=f(\pm\Lambda \pm \mu)$, provided the integral on the l.h.s.\ exists.

Let us now consider the scaling limit of the three consistency conditions~\eref{eq:cond}.
The $n=0$ constraint gives, after taking into account the boundary conditions (\ref{eq:Uscinf}, \ref{eq:Uscminf}), the trivial relation $\sum_a(-1)^a (N^+_{h,a}+N^-_{h,a})=0$, which  holds because
 $N^\pm_{h,a}=N^\pm_{h}$, see Sec.~\ref{sec:cf}.
The $n=1$ constraint yields, after applying eq.~\eref{eq:split}, the non-trivial identity $A_1^+=A_1^-$.
Also, using the asymptotic behaviour
\begin{equation}
  k_h(\lambda) \sim
  \pm \pi \lambda/(\pi-2\gamma) \,,\qquad
	\lambda \to \pm\infty
	\label{eq:kh_as}
\end{equation}
we get from~\eref{eq:S2}:
\begin{equation} \label{eq:S3}
  K = 2\gamma(A_1^+ + A_1^-)
  = 4\gamma A_1^\pm \,.
\end{equation}
Moreover, after substituting $\eta_{b\ell} \to \pm(\Lambda + \eta_{b\ell}^\pm)$ in
the consistency condition~\eref{eq:cond} for $n=2$, applying~\eref{eq:split} and expanding in powers of $\Lambda$, we obtain
the relation
\begin{equation*}
  2 \Lambda (A_1^+ + A_1^-) + (A_2^+ + A_2^-) = -\pi \wt m \,.
\end{equation*}
Inserting~\eref{eq:S3} and using the definition of $\Lambda$, we get
\begin{equation} 
  \wt m = -\frac{\pi-2\gamma}{\pi^2\gamma} \ K\log L
  -\frac{1}{\pi}(A_2^+ + A_2^-) \,. \label{eq:mt-s}
\end{equation}
This equation gives the precise scaling of the integer $\wt m$ as $L \to\infty$
with $K$ fixed.
Moreover, as it was previously argued in~\cite{Ikhlef:2011ay},
the constant term $(A_2^++A_2^-)$
is related to the finite part of the density of states in
the energy spectrum: see~\secref{density}.

Let us comment on the nature of the source terms
proportional to $(-1)^a$ in the NLIE~\eref{eq:NLIE-scaling}.
The first term equals $K \lambda/2\gamma$. Recall that,
in our setting, $K$ is a fixed parameter of the problem, exactly
like $m$, $e$ and $\vphi$, and so this term should be viewed as an input data of the NLIE.
In contrast, the value of the second term $-A_2^+$ is fixed by the boundary condition~\eref{eq:Zinf-shift} at 
$\lambda \to +\infty$. Our approach does not give an explicit
expression of $A_2^\pm$ in terms of $(m,e,K)$, but rather in terms of
the solution of the NLIE.


\section{Energy spectrum and higher spin charges}\label{sec:spec}

\subsection{Exact energy spectrum}
\label{sec:energy}


Using the asymptotic behaviour
\begin{equation}
  \epsilon_h(\lambda) \mathop{\sim}_{\lambda \to \pm\infty}
  2v_F \ e(\pm\lambda) \,,
\end{equation}
and eq.~\eref{eq:split}, the total energy~\eref{eq:E2} takes the form $E = 2Le_\infty + E^++E^-$, with
\begin{equation}
  E^\pm = \frac{2v_F}{L} \Big\{
  \sum_{b,j} \nu_{bj}^\pm e(\eta_{bj}^\pm)
   - \frac{1}{\pi} \sum_b
    \int_{\Omega_b} d\lambda\ \Im[e'(\lambda) U_b^\pm(\lambda)] \Big\} \,.
\end{equation}
For the sake of clarity, we shall first explain the calculation of $E^+$ and in the end list the modification required to compute $E^-$.

First, we shall restrict the NLIE~\eref{eq:NLIE-scaling} to the real axis. Let us define the functions:
\begin{equation} \label{eq:def-Q}
  Q^+_a(\Re\,\lambda) := \lim_{\delta \to 0}  2\Im\,
  U_a^+(\lambda) \,,
	\qquad
	\lambda\in\Omega_a\,,
\end{equation}
where the dependence on $\delta$ is contained in $\Omega_a$, and the 
even and odd combinations:
\begin{eqnarray*}
  Q^+_\even := Q^+_0 + Q^+_1 \,, \qquad
  Q^+_\odd := Q^+_0 - Q^+_1 \,.
\end{eqnarray*}
Then the NLIE eqs.~(\ref{eq:NLIE-ev-scaling}, \ref{eq:NLIE-odd-scaling}) take the compact form
\begin{eqnarray*}
Z^+_\even(\lambda)&=&-4 e(\lambda)+2C^+ - \sum_{b,l}\nu_{b\ell}^+ H_\even(\lambda-\eta_{b\ell}^+)-J_\even\star Q^+_\even(\lambda)\,,\\
Z^+_\odd(\lambda)&=&4A_1^+\lambda - 2 A_2^+ - \sum_{b,l}(-1)^b\nu_{b\ell}^+ \widetilde{H}_\odd(\lambda-\eta_{b\ell}^+)-\widetilde{J}_\odd\star Q^+_\odd(\lambda)\,.
\end{eqnarray*}
Notice that the derivative $(Q_a^+)'$ is not well-defined, and so it was important to integrate
by parts {\it before} taking the limit.
In the following it will be convenient to work with a more symmetric form of the above equations
\begin{eqnarray}\label{eq:suitNLIE}
Z^+_\even(\lambda)&=&-4 e(\lambda)+2C^+ - \sum_{b,l}\nu_{b\ell}^+ H_\even(\lambda-\eta_{b\ell}^+)-J_\even\star Q^+_\even(\lambda)\,,\\
Z^+_\odd(\lambda)&=&2A_1^+\lambda -  A_2^+ - \sum_{b,l}(-1)^b\nu_{b\ell}^+ H_\odd(\lambda-\eta_{b\ell}^+)-J_\odd\star Q^+_\odd(\lambda)\,,\nn
\end{eqnarray}
which is well defined because $Q^+_\odd$ decays exponentially at $+\infty$, see eq.~\eref{eq:Uscinf}, and where
\begin{equation*}
A_n^+ = \alpha \Big[\sum_{b,l}(-1)^b \nu_{b\ell}^+ - \frac{1}{2\pi}\int d\mu\, n\mu^{n-1} Q_\odd^+(\mu)\Big]\,.
\end{equation*}

Next, we manipulate in the standard way the NLIE eqs.~\eref{eq:suitNLIE} in order to compute
\begin{equation} \label{eq:Epm}
  E^+ = \frac{2v_F}{L} \Big[
  \sum_{b,j} \nu_{bj}^+ e(\eta_{bj}^+)
  - \frac{1}{2\pi} 
  \int d\lambda\ e'(\lambda) Q_\even^+(\lambda)
  \Big] \,.
\end{equation}
First step is to use the last equation in \eref{eq:NLIE-scaling}
\begin{equation*}
  \fl\qquad  2\pi I_h^+:=2\pi \sum_{a,j} \nu_{aj}^+ I_{h,aj}^+= \sum_{a,j} \nu_{aj}^+ Z_a^+(\eta_{aj}^+)\,,
\end{equation*}
in order to get after some simple algebra
\begin{eqnarray}\label{eq:sum-BAE}
  \fl\qquad2\pi I_h^+ &=& -2\sum_{b,j} \nu_{bj}^+ e(\eta_{bj}^+) + 2N_h^+ C^+
	+ A_1^+ \sum_{aj}(-1)^a \nu_{aj}^+ \eta_{aj}^+ \\
	\fl \quad&& - \half \sum_{aj}\nu_{aj}\int d\lambda [J_\even(\lambda-\eta_{aj}^+)Q_\even^+(\lambda)
+ (-1)^aJ_\odd(\lambda-\eta_{aj}^+)Q_\odd^+(\lambda)]
	\,.\nn
\end{eqnarray}
If we now use this expression to compute the sum over $\nu_{bj}^+ e(\eta_{bj}^+)$ and inject the result into~\eref{eq:Epm} then we get
\begin{eqnarray*}
  \fl\quad \frac{L E^+}{v_F} = -2\pi I_h^+ + 2N_h^+ C^+ +\frac{(A_1^+)^2}{\alpha} +\int \frac {d\lambda}{4\pi}\, Q_\even^+(\lambda)\Big[-4 e'(\lambda)-\sum_{a,j}\nu_{aj}^+H_\even'(\lambda-\eta_{aj}^+)\Big]\\
	\fl \qquad + \int \frac{d\lambda}{4\pi}\, Q_\odd^+(\lambda)\Big[2A_1^+ -\sum_{a,j}(-1)^a\nu^+_{aj}H_\odd'(\lambda-\eta_{aj}^+)\Big]
	\,.
\end{eqnarray*}
In the integrals, the factors in the brackets are the derivatives of the source terms
of the NLIE~\eref{eq:suitNLIE}.
Hence, we can write:
\begin{eqnarray}\label{eq:Eint}
  \fl \quad\frac{L E^+}{v_F} = -2\pi I_h^+ + 2N_h^+ C^+
  +\frac{(A_1^+)^2}{\alpha} 
  + \sum_a \int \frac{d\lambda}{2\pi} \ (Z_a^+)'(\lambda) \ Q_a^+(\lambda)  \\
  \fl \qquad \quad    +\sum_{a,b} \iint\frac{d\lambda\ d\mu}{4\pi}
  \left[Q_\even^+(\lambda) J_\even'(\lambda-\mu) Q_\even^+(\mu) +Q_\odd^+(\lambda) J_\odd'(\lambda-\mu) Q_\odd^+(\mu)\right]\,.\nn
\end{eqnarray}
After a change of variables $u=\exp(iZ^+_a)$, we obtain
\begin{equation*}
  \int \frac{d\lambda}{2\pi} \ (Z_a^+)'(\lambda) \ Q_a^+(\lambda)
  = -\frac{1}{\pi} \Re\ \left[
    \int_{\gamma_a^+} du\ \frac{\log(1+u)}{u}
    \right] \,,
\end{equation*}
where $\gamma_a^+$ is any path enclosed in the unit disk,
going from $0$ to $\exp[iZ_a^+(\infty)]=\exp[iQ_a^+(\infty)]$.
We choose the particular form of the integration path:
\begin{equation*}
  \int_{\gamma_a^+} du\ \frac{\log(1+u)}{u}
  = \int_0^1 du\ \frac{\log(1+u)}{u}
  + \int_0^{Q_a^+(\infty)} d(e^{i\theta}) \ \frac{\log(1+e^{i\theta})}{e^{i\theta}} \,.
\end{equation*}
Using formulas~(\ref{eq:dilog1}\,--\,\ref{eq:dilog2}), we get
\begin{equation} \label{eq:integral1}
  \int \frac{d\lambda}{2\pi} \ (Z_a^+)'(\lambda) \ Q_a^+(\lambda)
  = -\frac{\pi}{12} + \frac{[Q_a^+(\infty)]^2}{4\pi} \,.
\end{equation}
In the double integral of~\eref{eq:Eint} only the even part
contributes, because $J'_\odd$ is odd and $Q^+_\odd$ decays exponentially at $\pm\infty$. Applying the ``Lemma 1'' of~\cite{Destri:1997yz} we then get:
\begin{eqnarray} \label{eq:integral2}
  \iint\frac{d\lambda\ d\mu}{4\pi}
  Q_\even^+(\lambda) J_\even'(\lambda-\mu) Q_\even^+(\mu) &=& \frac{\pi-4\gamma}{32\pi\gamma} \ [Q_\even^+(\infty)]^2 \,.
\end{eqnarray}
Gathering the terms from~(\ref{eq:integral1}\,--\,\ref{eq:integral2}), we obtain
\begin{eqnarray}\label{eq:en_interm}
  \frac{L E^+}{v_F} &=& -2\pi I_h^+ + 2N_h^+ C^+ 
  +\frac{(A_1^+)^2}{\alpha} 
  -\frac{\pi}{6} + \frac{[Q_\even^+(\infty)]^2}{32\gamma} \,.
\end{eqnarray}
Finally, from eq.~\eref{eq:Uscinf} we get
\begin{equation*}
  Q_\even^+(\infty) = 4(\gamma m - \vphi) -4\pi(e+N_h^+) \,,
\end{equation*}
from eq.~\eref{eq:scaledBI} we can compute $I_h^+=N_h^+(N_h^++2e)$, and from eq.~\eref{eq:S2} we have $K=A^+_1/4\gamma$.
Combining everything together $N_h^+$ drops out of eq.~\eref{eq:en_interm} and we get:
\begin{eqnarray} \label{eq:E+}
  \frac{L E^+}{v_F} &=&  -\frac{\pi}{6}
  +\frac{\pi-2\gamma}{8\pi\gamma} K^2
  + \frac{1}{2\gamma} (\gamma m -\pi e+\vphi)^2 \,.
\end{eqnarray}

The calculation of $E^-$ is exactly analogous with the only difference being that all $+$-superscripts get replaced by $-$-superscripts and $e$, $\varphi$ flip signs
\begin{eqnarray} \label{eq:E-}
  \frac{L E^-}{v_F} &=&  -\frac{\pi}{6}
  +\frac{\pi-2\gamma}{8\pi\gamma} K^2
  + \frac{1}{2\gamma} (\gamma m +\pi e-\vphi)^2 \,.
\end{eqnarray}

\subsection{Identification to the SL(2,$\mathbb{R}$)/U(1)
sigma model}

From conformal invariance, we expect the following form for the scaling corrections to the energy and momentum:
\begin{eqnarray}\label{eq:exp_spec}
  \fl \qquad E^++E^- = \frac{2\pi v_F}{L} \left(
   h + \bar h -\frac{ c}{12}\right) \,,\qquad
	P =\frac{2\pi}{L} ( h - \bar h)\quad  \mod 2\pi \,, 
\end{eqnarray}
%
where $c$ is the central charge of the CFT and $h$, $\bar h$ are the conformal dimension of its primaries.
The total momentum $P$  is easily computed from eq.~\eref{eq:PE} by summing the BAE~\eref{eq:BAE2}:
\begin{equation}\label{eq:Pexpl}
  P = \frac{2\pi}{L} m \left(\frac{\vphi}{\pi} -e\right)\quad \mod 2\pi \,,
\end{equation}
which agrees with eq.~\eref{eq:mom}.
Comparing eqs.~(\ref{eq:exp_spec}) with eqs.~(\ref{eq:E+}, \ref{eq:E-}, \ref{eq:Pexpl}), we get:
\begin{equation} \label{eq:spectrum}
  \boxed{
    \qquad \begin{array}{l}
       \displaystyle h -\frac{c}{24}=-\frac{1}{12}
  +\frac{\pi-2\gamma}{16\pi^2\gamma} K^2
  + \frac{1}{4\pi\gamma} (\gamma m -\pi e+\vphi)^2 \,,\\ 
	\\[-10pt]
       \displaystyle \bar h -\frac{c}{24}=-\frac{1}{12}
  +\frac{\pi-2\gamma}{16\pi^2\gamma} K^2
  + \frac{1}{4\pi\gamma} (\gamma m +\pi e-\vphi)^2 \,.
  \end{array} \qquad}
\end{equation}
We emphasise the fact that this conformal spectrum was obtained analytically from the scaled NLIE of the lattice model\footnote{
Note that the dimensions~\eref{eq:spectrum} are obtained under the assumption that all Bethe roots $\lambda_{aj}$ are real. However, for large enough values of $\vphi$, a solution with complex roots exists,
and becomes the lowest energy state in the sector with $(e,m,\vphi)$ fixed~\cite{Jacobsen:2005xz}.
Although this type of state plays an important role in the statistical model, its study is beyond the
scope of the present paper.
}.

Notice that the spectrum~\eref{eq:spectrum} has a continuous quantum number --- the quasi-momentum $K$ --- corresponding to the quasi-shift conserved charge \eref{eq:light-cone} and its eigenvalue \eref{eq:S}.
Therefore, our spin chain  spectrum must correspond to a non-rational CFT with a non-normalizable Virasoro vacuum, i.e.\ we do not expect a state with $h=\bar{h}=0$ to be part of its spectrum.
Instead, the spin chain vacuum, characterized by $e=m=K=0$, should be identified with the primary state of the CFT
with the lowest possible conformal dimension.

If we now identify
\begin{equation}
s := \frac{\pi-2\gamma}{4\pi \gamma}\, K\,,\qquad  k := \frac{\pi}{\gamma} \,,
\label{eq:identf}
\end{equation}
where $k \in ]2,\infty[$, and set $\vphi=0$ 
then the spectrum~\eref{eq:spectrum}  coincides exactly with the continuous component of the
SL(2,$\mathbb{R})/$U(1) sigma model spectrum at level $k$ given by
\begin{eqnarray}
  c &=& \frac{3k}{k-2} -1 \,, \\
	 h &=& \frac{(m-ke)^2}{4k} + \frac{s^2+1/4}{k-2} \,, \\
  \bar h &=& \frac{(m+ke)^2}{4k} + \frac{s^2+1/4}{k-2} \,,
\end{eqnarray}
where we recall that  primaries of the SL$(2,\mathbb{R})$/U(1) coset algebra are labelled by pairs of SL$(2,\mathbb{R})$ affine primaries of spin $j= -1/2 + i s$ from the continuous series and U(1) vertex operators with ``winding number'' $e$ and ``momentum'' $m$.
Geometrically, $2s$ is interpreted as the momentum carried by the string in the non-compact axial direction of SL$(2,\mathbb{R})$/U(1), which is shaped as an infinite cigar, while $e$ and $m$ are the winding and momentum numbers in the compact direction.

Notice that the state with the lowest conformal dimension $h_0=\bar h_0 = 1/4(k-2)$ corresponds, as expected, to the ground state of the spin chain.
The effective central charge observed in the spin chain (within the pure hole sector) is then $c_{\rm eff}=c-24 h_0 = 2$, which is illustrative of the fact that there are two hole species.



\subsection{Scaling limit of transfer matrices}

In this section we shall take the scaling limit of the generating functions for the higher order conserved charges on the lattice using the expressions obtained in Sec.~\ref{sec:hs}.
The goal is to derive some equations that allow, at least in principle, to compute the mutually commuting local integrals of motion of the scaling theory in terms of the solution to the NLIE.
This type of equations serve to pin down the integrable structure of the scaling CFT, i.e.\ of the SL$(2,\mathbb{R})$/U(1) sigma model.
\subsubsection{Free energy}
First, let us show that if we scale the lattice transfer matrices according to eq.~\eref{eq:scaling} while keeping the spectral parameter $u$ finite, then the result can be expressed in terms of $E^\pm$ and $K$, i.e.\ one does not produce a generating function 
for the higher order conserved charges.

Indeed, notice that for $-\pi/4+\gamma/2=u_0<u<0$ one has
\begin{equation*}
  \fl \qquad |\Ub_a(2iu_0-2iu)|\propto \left[ \frac{-\sin 2u}{\sin 2(\gamma-u)} \right]^L
 \ll 1 \,,
\end{equation*}
and hence these terms vanish exponentially with $L$.
Now, taking the scaling limit of eq.~\eref{eq:fsol} with the help of eqs.~(\ref{eq:phlatt}, \ref{eq:asymptsig}) and (\ref{eq:split}, \ref{eq:Uscinf}),
then using $e(\lambda+2iu) = e(\lambda)e(2iu)$ and regrouping terms, we get
\begin{equation*}
  \fl\qquad F(u) \approx L f_\infty(u)
	- i \big[e(2iu) E^+ -{} e(-2iu) E^-\big]/v_F 
	\quad \mod \pi i\,,
\end{equation*}
up to exponentially small terms in $L$.
The first term is the bulk free energy that we can simply subtract.
The  scaling correction term becomes,  after using eqs.~(\ref{eq:E+}\,--\,\ref{eq:E-}),
\begin{equation*}
   \frac{1}{L} \left\{
    \frac{\pi c \sin 2\pi u/(\pi-2\gamma)}{12}  - 2i\pi\left[e(2iu) h - e(-2iu) \bar h\right]
    \right\} \,.
\end{equation*}
At the isotropic point $u\to u_0$, this coincides with the asymptotic behaviour
predicted by CFT. Similarly, taking the scaling limit of $G(u)$ in eq.~\eref{eq:gsol} using eq.~\eref{eq:kh_as} we get
\begin{equation*}
G(u)\approx K\,,
\end{equation*}
up to exponentially small terms in $L$.

\subsubsection{Higher spin  charges} Let us now take a different scaling limit of $F(u)$, $G(u)$
\begin{eqnarray}\nn
F^\pm(u)&:=&\lim_{L\to\infty}\pm  \left[F(\pm u\pm i\Lambda/2) - 2L f_\infty(\pm u\pm i\Lambda/2)\right]\\
G^\pm(u)&:=& \lim_{L\to\infty} G(\pm u\pm i\Lambda/2)\,.
\label{eq:defFUsc}
\end{eqnarray}
Then, a straightforward calculation leads to
\begin{eqnarray}\nn
\fl \quad F^+(u)=-i \sum_{b,\ell} \nu_{b\ell}^+\, \tilde p_h(\eta_{b\ell}^++2iu) - (\Ub^+_0+\Ub^+_1)(2iu_0-2iu)&\\ \nn
  \fl\qquad {}- \frac{1}{2\pi} \sum_b   \left[
    \int_{\Omega_b} d\lambda \ \tilde p_h(\lambda+2iu) (U^+_b)'(\lambda)
    -\int_{\overline{\Omega}_b} d\lambda \tilde p_h(\lambda+2iu) (\overline{U}^+_b)'(\lambda)
    \right]\,,\\
		\fl \quad F^-(u)=-i\sum_{b,\ell}\nu_{b\ell}^-\, \tilde p_h(\eta_{b\ell}^-+2iu)+ (U^-_0+U^-_1)(-2iu-2iu_0)&\\ \nn
  \fl\qquad {}- \frac{1}{2\pi} \sum_b \left[
    \int_{\Omega_b} d\lambda \ \tilde p_h(\lambda+2iu) (U^-_b)'(\lambda)
    -\int_{\overline{\Omega}_b} d\lambda \ \tilde p_h(\lambda+2iu) (\overline{U}^-_b)'(\lambda)
    \right]\,
\label{eq:scF}
\end{eqnarray}
modulo $2\pi$ and, similarly,
\begin{eqnarray}\nn
\fl \quad G^+(u)=K+ \sum_{b,\ell}(-1)^b \nu_{b\ell}^+\, \tilde k_h(\eta_{b\ell}^++2iu)+ (\Ub^+_0-\Ub^+_1)(2iu_0-2iu)&\\ \nn
  \fl\qquad {}+ \frac{1}{2i\pi} \sum_b (-1)^b  \left[
    \int_{\Omega_b} d\lambda \ \tilde k_h(\lambda+2iu) (U^+_b)'(\lambda)
    -\int_{\overline{\Omega}_b} d\lambda \tilde k_h(\lambda+2iu) (\overline{U}^+_b)'(\lambda)
    \right]\,,\\
		\fl \quad G^-(u)=K+ \sum_{b,\ell}(-1)^b \nu_{b\ell}^-\, \tilde k_h(\eta_{b\ell}^- +2iu)+ (U^-_0-U^-_1)(-2iu-2iu_0)&\\ \nn
  \fl\qquad {}+ \frac{1}{2i\pi} \sum_b (-1)^b  \left[
    \int_{\Omega_b} d\lambda \ \tilde k_h(\lambda+2iu) (U^-_b)'(\lambda)
    -\int_{\overline{\Omega}_b} d\lambda \ \tilde k_h(\lambda+2iu) (\overline{U}^-_b)'(\lambda)
    \right]\,.
\label{eq:scG}
\end{eqnarray}
Here we have defined the functions
\begin{eqnarray*}
  \tilde p_h(\lambda) &=& p_h(\lambda)+\pi/2 =  2 \sum_{n=0}^\infty \frac{(-1)^n}{2n+1} e(\lambda)^{2n+1} \,, \\
  \tilde k_h(\lambda) &=& k_h(\lambda) - \pi\lambda/(\pi-2\gamma)=   -\sum_{n=1}^\infty \frac{(-1)^n}{n} \ e(\lambda)^{2n} \,
\end{eqnarray*}
 decaying exponentially at $\lambda\to+\infty$.

First, notice that $F^\pm$ and $G^\pm$ have a form very similar to their lattice counterparts (\ref{eq:fsol}, \ref{eq:gsol}). Therefore, it is reasonable to expect that
they satisfy a Baxter type equation in analogy with eq.~\eref{eq:Lambda}.

Secondly, expanding $F^\pm(u)$, $G^\pm(u)$ around $\Re\, 2iu\to +\infty\pm i \delta$, where $\Ub^+_a$ and, respectively, $U^-_a$ have a faster then exponential decay, see eq.~\eref{eq:Uscminf}
and the source terms of eq.~\eref{eq:NLIE-scaling},
we get an asymptotic expansion in powers of $e(2iu)$
\begin{eqnarray}
F^\pm(u)\simeq \sum_{n=0}^\infty e(2iu)^{2n+1}\, F^\pm_{2n+1}\,,\qquad G^\pm(u)\simeq \sum_{n=0}^\infty e(2iu)^{2n}\, G^\pm_{2n}\,.
\label{eq:hs_exp}
\end{eqnarray}
The dominant terms of this expansion are given by
\begin{eqnarray*}
F^+_1 =-2\pi i\,  (h-c/24)\,,\qquad F^-_1 =-2\pi i\,  (\bar{h}-c/24)\,, \qquad G^\pm_0 = K\,.
\end{eqnarray*}
Eqs.~\eref{eq:hs_exp} generate the entire hierarchy of mutually commuting local conserved charges of the scaling CFT, i.e.\ the SL(2,$\mathbb{R}$)/U(1) sigma model.
With this interpretation $e(2iu)$  has the natural scale of [energy]$^{-1}$, while $F^\pm_{2n-1}$ are  the conserved charges of  holomorphic currents of spin $2n$ and, finally,
$G^\pm_{2n}$ are  the conserved charges of  holomorphic currents of spin $2n+1$.
This is in agreement with the expectation \cite{Schiff:1992tv} that the integrable structure of the black hole CFT is described by the  non-linear Schr\"odinger equation.
We leave the comparison of the higher spin charges following from eqs.~(\ref{eq:scF}, \ref{eq:scF}) with the CFT predictions of \cite{Schiff:1992tv, Fateev:1990bf, Bakas:1991fs, Yu:1991cv} to future work.


\section{Density of states}\label{sec:dens}

\subsection{Spin chain and CFT definitions}
\label{sec:density}

The matching of spectra suggests that the staggered
six-vertex model has a conformal scaling limit described  by the
SL(2,$\mathbb{R}$)/U(1) sigma model. However, since the 
spectra are continuous, the identification between the string axial momentum $2s$ and the quasi-momentum $K$ given in eq.~\eref{eq:identf} is not very convincing --- the matching would work for any positive constant multiplying $K^2$ in eq.~\eref{eq:spectrum}.
To eliminate this ambiguity, the identification \eref{eq:identf} must be accompanied by a comparison of the densities of states.

The density of states in the spin chain is a quantity diverging with the system size $L$ and defined as follows.
First, let us fix the quantum numbers $(m,e)$.
Then, from~\eref{eq:mt-s} we have
\begin{equation}
  \wt m = -\frac{4s}{\pi} \left[
    \log \frac{L}{L_0} + B
    \right] \,,
\end{equation}
where 
\begin{equation} \label{eq:B-A2}
  B(s) := \frac{1}{4s} (A_2^+ + A_2^-) + \log L_0 \,,
\end{equation}
and $L_0$ is an arbitrary constant, depending only on $\gamma$, which explicitly takes into account the ambiguity of separating the logarithmically divergent term from the finite part $B(s)$.
The allowed values of $\wt m$ are all the positive integers with the same
parity as $m$. Hence, they differ by steps of $\delta\wt m=2$.
If one increases $\wt m$ by a finite amount, one gets\begin{equation}
  \delta\wt m = \frac{4\delta s}{\pi} \left[\log \frac{L}{L_0} + \partial_s(sB) \right] \,,
\end{equation}
and any sum over $\wt m$ becomes, in the scaling limit:
\begin{equation}
  \sum_{\wt m}  (\dots)
  \longrightarrow \int ds \ \rho(s) \ (\dots) \,,
\end{equation}
where the density of states is
\begin{equation}
  \rho(s) := \frac{\delta\wt m}{2\delta s}
  = \frac{2}{\pi} \left[\log \frac{L}{L_0} + \partial_s(sB) \right] \,.
\end{equation}
Thus, we have found that the source terms $A_2^\pm$ are directly related
to the finite part of the density of states for the scaling limit
of the staggered six-vertex model.

There is a heuristic way to compute the large-$s$ behaviour of $B(s)$.
The source term of
the NLIE~\eref{eq:NLIE-scaling} for $Z^\pm_0$ is up to exponentially small terms given by
$\sigma_0(s):=-2e(\lambda)+ C^\pm +2\pi s\lambda/(\pi-2\gamma)-A_2^\pm$.
It has a maximum at $\lambda^* = -(1-2\gamma/\pi) \log(-s)$,
with $\sigma_0(\lambda^*)=-2s[\log(-s)-1]+C^\pm-A_2^\pm$.
From eqs.~(\ref{eq:mt-s}, \ref{eq:identf}) it follows that $s<0$ since $\wt m>0$.
For large $-s$, this
maximum has a large value, and the number of extraordinary pairs of holes
becomes large, too. If we approximate $Z^\pm_0(\lambda)$ by $\sigma_0(\lambda)$ then it is easy to  see that for large $s$ the maximum becomes highly peaked
and, thus, one can assume that the majority of holes are closely and centrally distributed around $\lambda^*$
\begin{equation}
  \{ \eta_{0j}^+ \} \simeq \{ \lambda^* \pm \delta\eta_{0j}^+ \}\,,\qquad \{ \eta_{0j}^- \} \simeq \{ \lambda^* \pm \delta\eta_{0j}^- \} \,.
\end{equation}
In this approximation, we have, from~\eref{eq:def-An}:
\begin{equation}
  A_1^\pm \simeq -2\alpha \sum_j \delta\eta_{0j}^\pm \,,
  \qquad
  A_2^\pm \simeq -4\alpha \lambda^* \sum_j \delta\eta_{0j}^\pm \,,
\end{equation}
where the  contribution of integrals  and ordinary holes is subdominant.
Thus we get $A_2^\pm \simeq -2s \log(-s)$, which then gives from eq.~\eref{eq:B-A2} the asymptotic
\begin{equation}
  B(s) \simeq -\log(-s) \qquad \text{for $s\to -\infty$.}
\end{equation}

In comparison, the density of states in the SL(2,$\mathbb{R}$)/U(1)
sigma model has the form
\begin{eqnarray}\label{eq:BH_dens}
  \rho_{\rm BH}(s) 
  &=& \frac{1}{\pi} \left[\log \epsilon + \partial_s(sB_{\rm BH}) \right] \,, \\ \nn
  B_{\rm BH}(s) &=& \frac{1}{2s} \Im\log \left[
    \Gamma\left( \frac{1-m+ek}{2} - is \right)
    \Gamma\left( \frac{1-m-ek}{2} - is \right)
    \right] \,,
\end{eqnarray}
see \cite{Hanany:2002ev}.
It is useful to recall that this density of states was also computed by discretizing the spectrum of the axial string momentum $2s$ by adding a Liouville wall to the action, which  confines
the movement of the centre of the string in the axial direction to a region of length $\log \epsilon$. The finite part of the density of states
can then be extracted from the reflection amplitudes of the SL$(2,\mathbb{R})$/U(1) sigma model at the tip of the cigar and of the Liouville theory off of the Liouville wall.
More precisely, $8sB_{\mathrm{BH}}$ is the difference of the two reflection amplitudes, respectively.

Notice that our function $B(s)$
has the correct behaviour at $s \to -\infty$. For finite $s$, the
expression~\eref{eq:def-An} for $A_2^\pm$ does not lend itself to an analytical
computation like in~\secref{energy}, essentially because $A_2^\pm$ are not local conserved
quantities of the Hamiltonian. Thus, in~\secref{num}, we shall study the NLIE
numerically to obtain the full function $B(s)$.

\subsection{Numerical comparison}
\label{sec:num}

\subsubsection{Numerical algorithm}

The main purposes of our algorithm are (i) to check the validity of
the scaling regime by providing a numerical solution of the
NLIE~\eref{eq:NLIE-scaling} and (ii) to compute the integration
constants $A_2^\pm$, which give access to the finite part of
the density of states.

In this section, although we work with the scaled
NLIE~\eref{eq:NLIE-scaling}, we omit the $\pm$ indices to
lighten the notation. The system we have to solve is
\begin{eqnarray}
  \fl\qquad Z_a(\lambda) &=& \sigma_a(\lambda)
  - \sum_{b,\ell} \nu_{b\ell} \wt H_{a-b}(\lambda-\eta_{b\ell}) \nn \\
  \fl\qquad && - \frac{1}{i} \sum_b \left[
    \int_{\Omega_b} d\mu \ \wt J_{a-b}(\lambda-\mu) U_b(\mu)
    - \int_{\overline{\Omega}_b} d\mu \ \wt J_{a-b}(\lambda-\mu) \Ub_b(\mu)
    \right] \,, \label{eq:NLIE-num} \\
  \fl\qquad Z_a(\eta_{aj}) &=& 2\pi I_{h,aj} \,, \label{eq:BAE-num}
\end{eqnarray}
where the full source term is given by
\begin{equation}
  \fl\qquad \sigma_a(\lambda) := -2e(\lambda) + C +
  (-1)^a \left( \frac{2\pi s}{\pi-2\gamma} \ \lambda - A_2 \right) \,.
\end{equation}
In~\eref{eq:NLIE-num}, the number of ordinary holes $N_h$ is given by~\eref{eq:Nh},
while the integration constant $C$ by \eref{eq:NLIE-scaling}. In contrast,
the constant $A_2$ is not known, and it determines, among other things, the number of
extraordinary holes $\wt N_h$. As explained in~\secref{source-terms}, the correct
value of $A_2$ is the one for which both $Z_0(\lambda)$ and $Z_1(\lambda)$
converge to their expected limit~\eref{eq:Zinf-shift} at $\lambda \to +\infty$.
Thus, our algorithm uses a trial value for $A_2$, and evolves it to get
the $Z_a(+\infty)$ as close as possible to their expected values.

Let us now explain how the algorithm solves the system~(\ref{eq:NLIE-num}\,--\,\ref{eq:BAE-num})
for a given value of $A_2$. First, we discretise  the paths $\Omega_a$, and write
the integrals as
\begin{equation}
  \fl\qquad \int_{\Omega_a} f(\mu) d\mu \longrightarrow
  \sum_{j=1}^N w_{aj} \ f(\mu_{aj}) \,,
\end{equation}
where $\{\mu_{aj}, w_{aj}\}$ are suitable points and weights for the approximation
of integrals over $\Omega_a$. Equation~\eref{eq:NLIE-num}
is then used in two different ways. First, equations~\eref{eq:NLIE-num}
for $\lambda = \mu_{a1}, \dots, \mu_{aN}$ form a closed system for the unknowns
$\{z_{aj}:=Z_a(\mu_{aj})\}$. Second, for $\lambda=\eta_{aj}$, the left-hand side
of~\eref{eq:NLIE-num} is replaced
by $2\pi I_{h,aj}$, and we have a system of BAE for the unknowns $\{\eta_{aj}\}$.
We denote by $\{ z_{aj}^{(0)}, z_{aj}^{(1)}, z_{aj}^{(2)} \dots \}$
the sequence of numerical estimates for $z_{aj}$, and
$u_a^{(n)}:= \log[1+\exp(iz_{aj}^{(n)})]$. Also, we have the sequence
of estimates $\{ \eta_{aj}^{(0)}, \eta_{aj}^{(1)}, \eta_{aj}^{(2)} \dots \}$
for $\eta_{aj}$. We define the basic iteration, giving
$\{z_{aj}^{(n+1)}\}$ in terms of $\{u_{aj}^{(n)}\}$ and $\{\eta_{aj}^{(n)}\}$:
\begin{eqnarray}
  \fl\qquad z_{aj}^{(n+1)} &=& \sigma_a(\mu_{aj})
  - \sum_{b,\ell} \nu_{b\ell} \wt H_{a-b} \left(\mu_{aj}-\eta_{b\ell}^{(n)} \right) \nn \\
  \fl\qquad && - \frac{1}{i} \sum_b 
  \sum_{\ell=1}^N \left[
    w_{b\ell} \ \wt J_{a-b}(\mu_{aj}-\mu_{b\ell}) \ u_{b\ell}^{(n)}
    - \overline w_{b\ell} \ \wt J_{a-b}(\mu_{aj}-\overline\mu_{b\ell}) \ \overline u_{b\ell}^{(n)}
    \right] \,.
  \label{eq:NLIE-iter}
\end{eqnarray}
The $\{\eta_{aj}^{(n+1)}\}$ are found by solving the following non-linear system
by the multivariate Newton-Raphson method:
\begin{eqnarray}
  \fl\qquad 2\pi I_{h,aj} &=& \sigma_a \left(\eta_{aj}^{(n+1)} \right)
  - \sum_{b,\ell} \nu_{b\ell} \wt H_{a-b} \left(\eta_{aj}^{(n+1)}-\eta_{b\ell}^{(n+1)} \right) \nn \\
  \fl\qquad &&  - 2 \sum_b 
  \sum_{\ell=1}^N \Im \left[
    w_{b\ell} \ \wt J_{a-b} \left(\eta_{aj}^{(n+1)}-\mu_{b\ell} \right)\, u_{b\ell}^{(n)} \right] \,.
  \label{eq:BAE-iter}
\end{eqnarray}
The initial values in the algorithm are given by the source terms only
\begin{equation*}
\fl\qquad z_{aj}^{(0)}=\sigma(\mu_{aj})\,,\qquad 2\pi I_{h,aj}=\sigma(\eta_{aj})-\sum_{b,\ell}\nu_{b\ell}\wt H_{a-b}(\eta_{aj}-\eta_{b\ell})\,.
\end{equation*}
Moreover, at any point in the algorithm, we can evaluate $Z_a$ for real $\lambda$
by the extrapolation formula:
\begin{equation*}
  \fl\qquad Z_a^{(n)}(\lambda) := \sigma_a(\lambda)
  - \sum_{b,\ell} \nu_{b\ell} \wt H_{a-b} \left(\lambda-\eta_{b\ell}^{(n)} \right)
  - 2 \sum_b 
  \sum_{\ell=1}^N \Im \left[
    w_{b\ell} \ \wt J_{a-b}(\lambda-\mu_{b\ell}) \ u_{b\ell}^{(n)}
    \right] \,.
\end{equation*}
After enough iterations, we have reached, up to machine precision,
a fixed point for the counting functions
\begin{equation*}
  \fl\qquad Z_a^{(\infty)}:=\mathop{\lim}_{n\to \infty} Z_a^{(n)} \,.
\end{equation*}

\subsubsection{Numerical results}

For small values
of $A_2$, one has $\left(Z_0^{(\infty)}(+\infty),Z_1^{(\infty)}(+\infty) \right) = (-\infty,+\infty)$,
whereas for large values of $A_2$,
it is $(+\infty,-\infty)$. These two regimes are separated by a value $A_2^*$, for which
we observe $Z_0^{(\infty)}(+\infty)=Z_1^{(\infty)}(+\infty)=2\gamma m-2\vphi$.
We can follow a simple dichotomy procedure to find $A_2^*$.

Figures~\ref{fig:Z-num1}\,--\,\ref{fig:Z-num3} show the behaviour of $Z_0^{(n)}$ for various
values of $A_2$. \figref{B} shows the finite part of the density of states obtained
from the numerical solution of the NLIE, which displays excellent agreement
with the density of states in the SL(2,$\mathbb{R}$)/U(1) sigma model.


%
\begin{myfig}
  \includegraphics[scale=1]{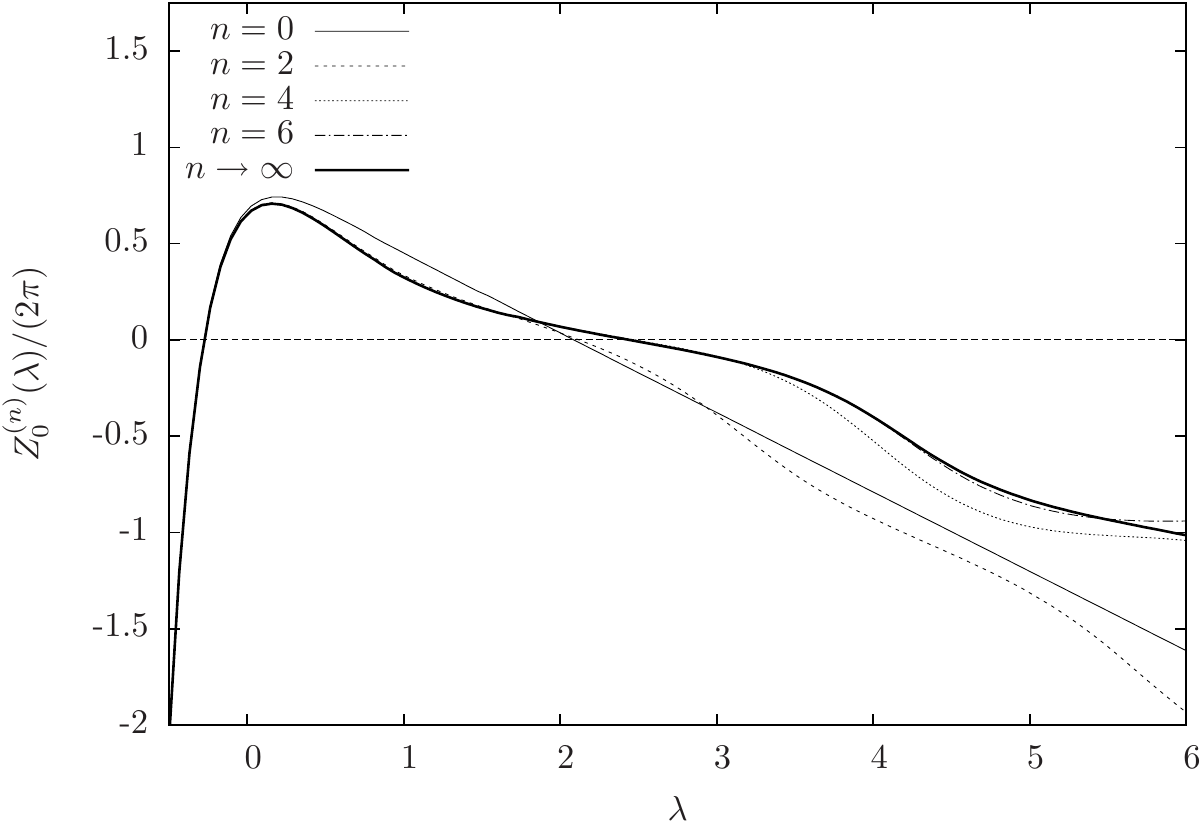}
  \caption{The iterates $Z_0^{(n)}$ for $\gamma=1.24$, $\vphi=0$, $m=e=0$, $s=-0.555$, when $A_2<A_2^*$.}
  \label{fig:Z-num1}
\end{myfig}
\begin{myfig}
  \includegraphics[scale=1]{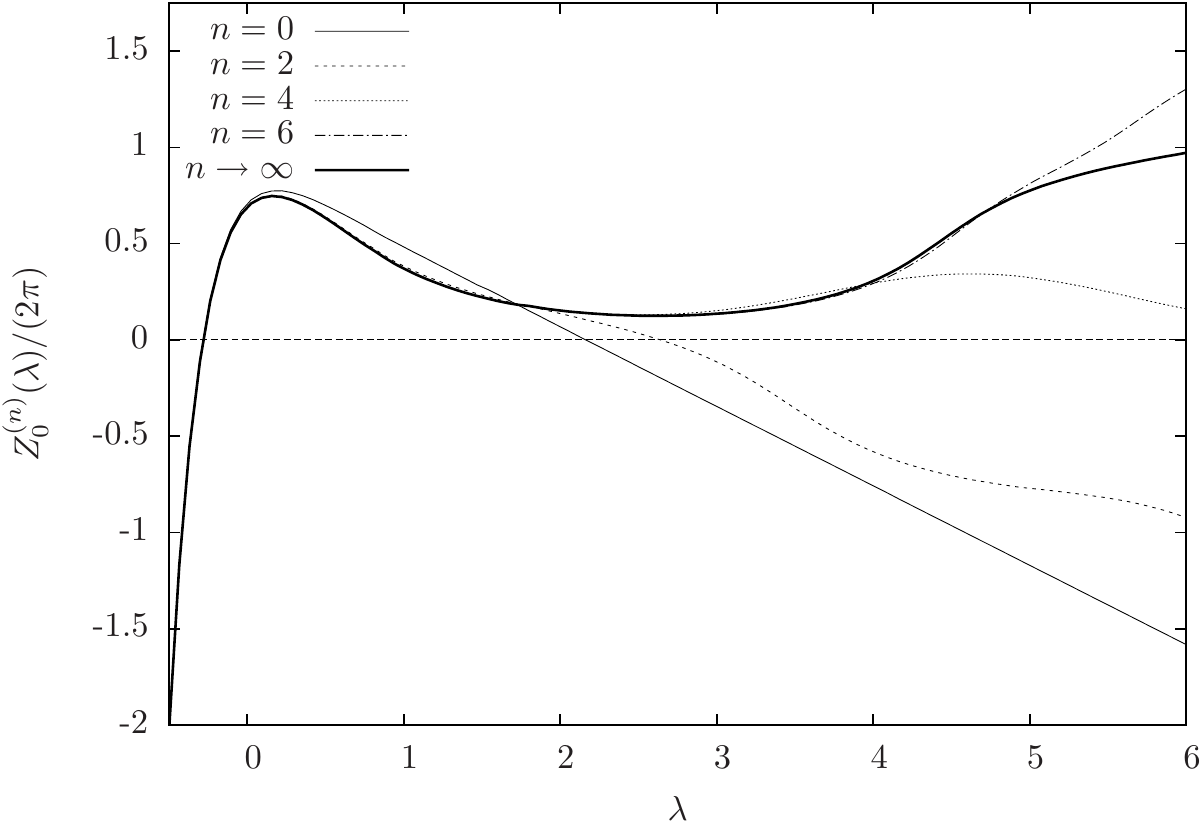}
  \caption{The iterates $Z_0^{(n)}$ for $\gamma=1.24$, $\vphi=0$, $m=e=0$, $s=-0.555$, when $A_2>A_2^*$.}
  \label{fig:Z-num2}
\end{myfig}
\begin{myfig}
  \includegraphics[scale=1]{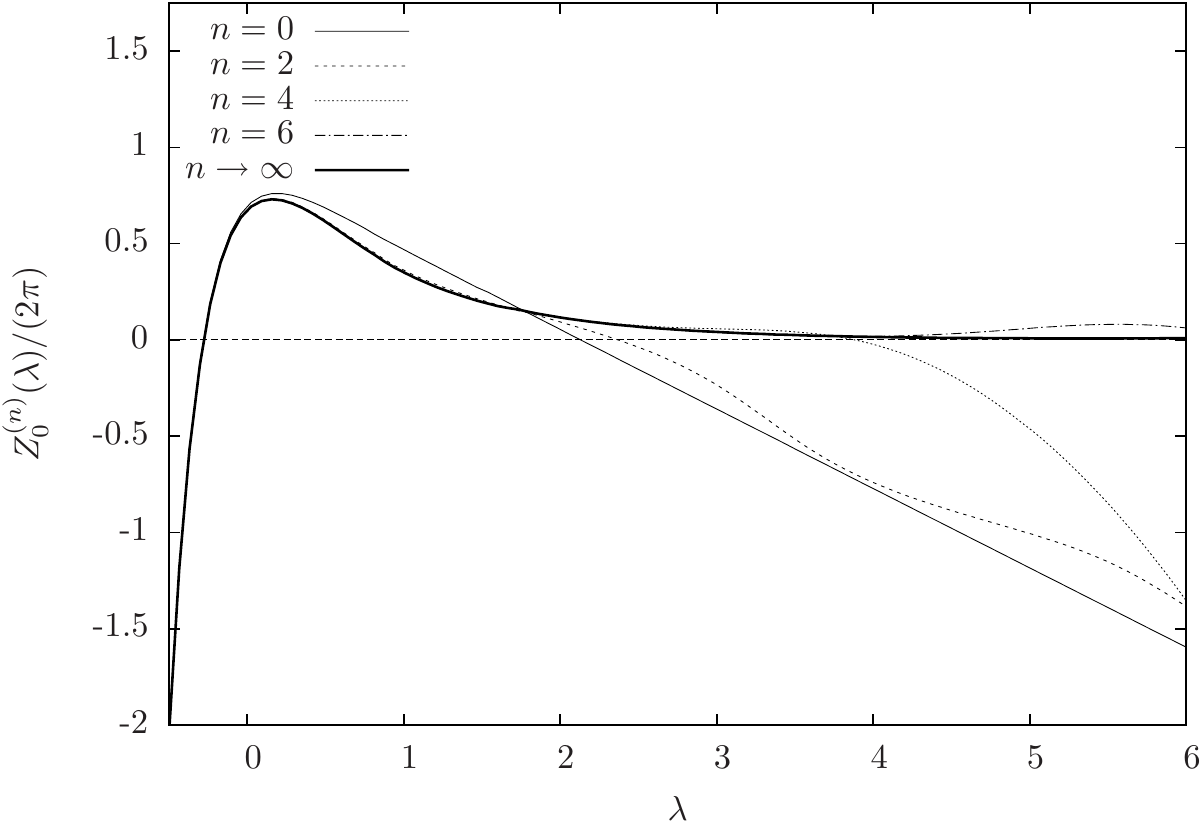}
  \caption{The iterates $Z_0^{(n)}$ for $\gamma=1.24$, $\vphi=0$, $m=e=0$, $s=-0.555$, when $A_2=A_2^*$.}
  \label{fig:Z-num3}
\end{myfig}
\begin{myfig}
  \includegraphics[scale=1]{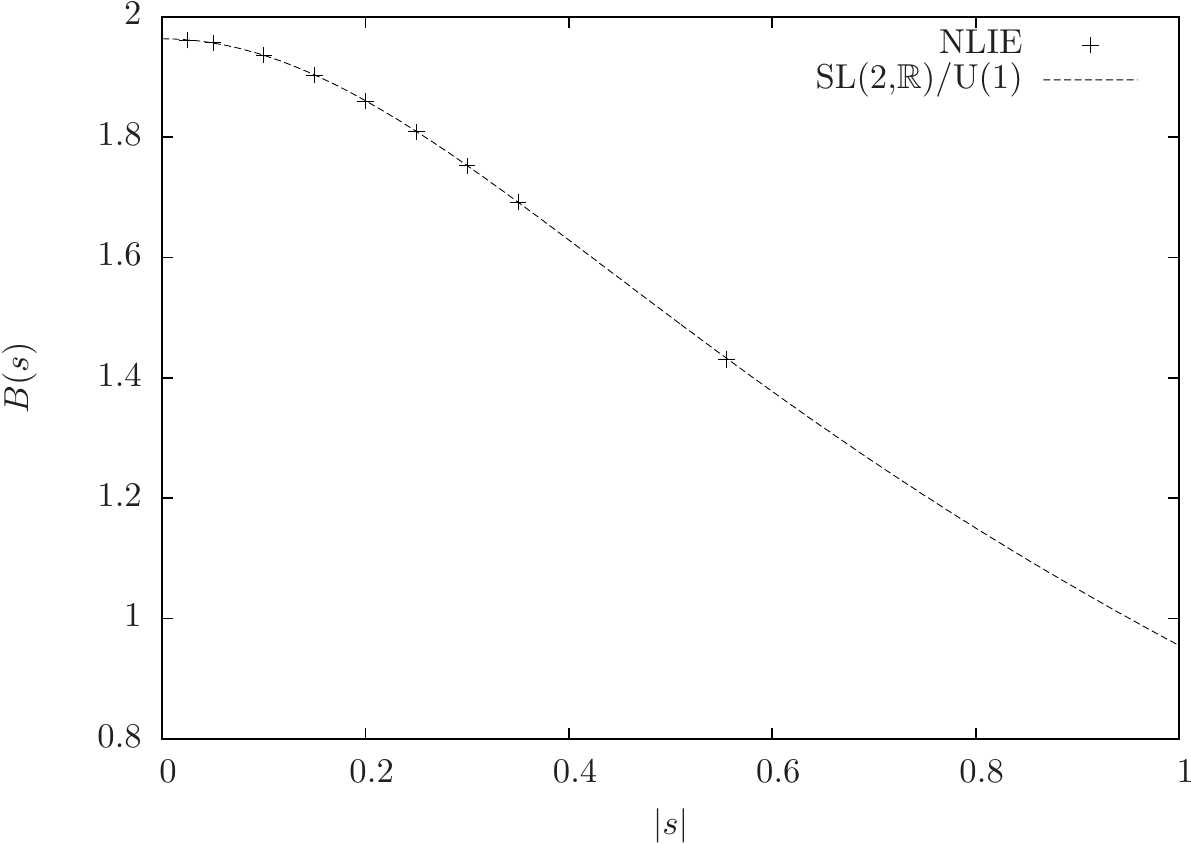}
  \caption{Finite part $B(s)$ of the density of states. Points are values
  obtained by solving numerically the NLIE and using~\eref{eq:B-A2}. The line
  is the function $B_{\mathrm{BH}}(s)$ calculated from the SL(2,$\mathbb{R}$)/U(1)
  WZW model, see eq.~\eref{eq:BH_dens}.}
  \label{fig:B}
\end{myfig}



\section{Discussion}


In this work we have considered the continuum limit of the critical staggered XXZ spin chain  defined in \cite{Jacobsen:2005xz} and further studied in \cite{Ikhlef:2008zz, Ikhlef:2011ay}.
Using the method of NLIEs to compute scaling corrections we have recovered the continuous spectrum computed in \cite{Ikhlef:2011ay} in the Wiener-Hopf approximation, which coincides with the continuous spectrum of the SL$(2,\mathbb{R})/$U(1) Euclidean black hole CFT.\footnote{We, however, did not reproduce the discrete spectrum of the black hole CFT, see \cite{Hanany:2002ev}.}
Additionally, we have numerically computed  with the NLIEs the density of states of the spin chain and found perfect agreement with the density of states of the black hole CFT.
The  NLIEs that we derived from the lattice displayed essentially new features such as: integral kernels that  do not decay at infinity,  non-monotonic solutions (i.e.\ counting functions) even in the absence of holes and unusual source terms/asymptotic behaviour in the
 region where the Bethe roots condense. Our analysis shows that these are closely related to the non-rational nature of the CFT.

We have only very briefly discussed the integrable structure of the black hole CFT and found that there is one conserved charge at every integer non-negative spin.
This is in agreement with the expectation of \cite{Schiff:1992tv} that the higher spin integrals of motions belong to the non-linear Schr\"odinger hierarchy.
It would be very interesting to study in more detail the integrable structure of the black hole CFT along the lines of 
 \cite{Bazhanov:1994ft, Bazhanov:1996dr, Bazhanov:1998dq} (see also \cite{Bytsko:2009mg}), i.e.\ construct the $\mathcal{Q}$-functions, the Baxter equation, the $T$-system, compute a few local and non-local higher spin integrals of motion, compare them to the CFT predictions of \cite{Schiff:1992tv, Fateev:1990bf, Bakas:1991fs, Yu:1991cv}
and find some ordinary differential equation reproducing these quantities via the ODE/IM correspondence of
\cite{Dorey:2007zx}.
Such a correspondence should make it possible to compute analytically the constants $A_2^\pm$ which appear in the NLIEs \eref{eq:NLIE-scaling} and which determine the density of states, see \cite{Fateev:2005kx, Teschner:2007ng, Bytsko:2009mg} for examples.
A good starting point is the ODE/IM correspondence of \cite{Lukyanov:2013wra}  for the Fateev SS  model,
which in a certain limit gives the black hole CFT.

Another interesting direction of research is to engineer a gap in the critical staggered XXZ spin chain following the standard
recipe of \cite{Destri:1987ug, Reshetikhin:1993wm} and take the continuum limit in such a way that the resulting NLIE describe an integrable massive perturbation of the black hole CFT.
The integrable structure  should remain invariant under the perturbation, which is clear on the lattice. We notice that there are at least two different integrable massive perturbations of the black hole CFT, known as complex Sinh-Gordon (CShG) models \cite{Pohlmeyer:1975nb, Lund:1976ze, Sciuto:1979ta, Getmanov:1980cq}, but only one of them has a spin 2 integral of motion \cite{Fateev:1990bf, Schiff:1992tv, Fateev:1996em} required by the lattice discretization.\footnote[1]{We thank Hubert Saleur for pointing this fact to us.}
The respective CShG model is classically defined by the action
\begin{equation*}
\mathcal{A} =\int  \frac{k}{2}\left[\frac{\partial_\mu\chi \partial_\mu\bar{\chi} }{1+\chi\bar\chi} - m^2 \chi\bar\chi\right]d^2 x\ ,
\end{equation*}
where we can recognize in the first term the original cigar metric~\eref{eq:cig_metr} in the complex coordinates $\chi=e^{i\varphi}\sinh \rho$ and $m^2$ is the coupling
to the massive perturbation.
This model is  classically integrable \cite{Pohlmeyer:1975nb, Lund:1976ze} and there are strong perturbative \cite{deVega:1981ka, deVega:1982sh, Bonneau:1984pj} and non-perturbative arguments \cite{Fateev:1996em, Fateev:1995ht, Ridout:2011wx} that it is also quantum integrable.
Its particle spectrum and exact $S$-matrix have been conjectured in \cite{Fateev:1995ht, Fateev:1996em, Dorey:1994mg}.
It would be very interesting to make sense of the scattering theory for the massive deformation of the staggered XXZ spin chain and see how it compares with the complex Sinh-Gordon model.
Partial results in this direction were obtained in \cite{HS}.

Finally, let us mention possible extensions of our work to supergroup spin chains, which arise naturally in the study of two-dimensional disordered quantum phase transitions, like the Integer Quantum Hall Effect~\cite{Zirnbauer:1999} or Spin Quantum Hall Effect~\cite{GLR:1999}. 
Under some specific conditions, some of these supergroup spin chains
are suspected~\cite{Essler:2005ag, Candu:2010qj, Frahm:2010ei, Frahm:2012eb} to have a continuous conformal spectrum in the scaling limit.
In the density approximation, they are characterized by the appearance of singular kernels in the linear integral form of the BAE, just like for the staggered XXZ spin chain, which produces a strongly degenerate spectrum.
It would be interesting to generalize the method of NLIE to these chains as well,
prove in this way the emergence of a continuous spectrum and compute its form together with the density of states. These data should allow to unambiguously identify the scaling CFTs, and ultimately to describe the
non-rational CFTs associated to some disordered quantum phase transitions.


\ack{We thank Sergei Lukyanov and  Hubert Saleur  for motivating and insightful discussions and J\"org Teschner for useful remarks on the manuscript.}

\appendix
\section[]{Useful formulas}

\paragraph{Proof of the summation formula~\eref{eq:sum}.} The contour
$C_a:= (-\Gamma_a) \cup \Gammab_a$ encloses all Bethe roots $\lambda$
counter-clockwise, and holes with $\nu_{aj}=+1$ (resp. $\nu_{aj}=-1$)
counter-clockwise (resp. clockwise). Since we chose $\delta$ in such a way that the roots and holes are the
only solutions to $1+(-1)^{r_a}e^{iZ_a}=0$ in the strip $|\Im\,\lambda|<\delta$,
we can write
\begin{equation*}
  \frac{1}{2i\pi} \oint_{C_a} \frac{(-1)^{r_a} i Z'_a(\lambda) e^{iZ_a(\lambda)}}
  {1+(-1)^{r_a} e^{iZ_a(\lambda)}} f(\lambda) \ d\lambda
    = \sum_j f(\lambda_{aj}) + \sum_j \nu_{aj} f(\eta_{aj}) \,.
\end{equation*}
We then substitute under the integral:
\begin{equation*}
  \frac{(-1)^{r_a} i Z'_a(\lambda) e^{iZ_a(\lambda)}}
  {1+(-1)^{r_a} e^{iZ_a(\lambda)}} = \left\{\begin{array}{ll}
  \frac{d}{d\lambda} \log[1+(-1)^{r_a} e^{iZ_a}]
  \quad &\lambda \in \Gamma_a \,, \\
  \frac{d}{d\lambda} \{ \log[1+(-1)^{r_a} e^{-iZ_a}] + iZ_a \}
  \quad &\lambda \in \Gammab_a \,,
  \end{array}\right.
\end{equation*}
which gives the relation~\eref{eq:sum}.
Finally, from the asymptotic of $U_a'(\lambda)\sim e^{-|\lambda|}$ at $\lambda\to\pm\infty$, which follows directly from eqs.~(\ref{eq:def-Z}, \ref{eq:Zinf}), we see that the integral in eq.~\eref{eq:sum} is well defined if $f(\lambda)$ grows slower then
$e^{a|\lambda|}$ with $a<1$ when $\lambda\to\pm\infty$.

\paragraph{Fourier transforms and convolution products:}
\begin{equation} \label{eq:FT}
  \wh{f}(\omega) := \int d\lambda \ f(\lambda) e^{i\omega\lambda} \,,
  \qquad
  f(\lambda) = \frac{1}{2\pi} \int d\omega \ \wh{f}(\omega) e^{-i\omega\lambda} \,.
\end{equation}
\begin{equation} \label{eq:convol}
  (f \star g)(\lambda) := \int d\mu \ f(\mu)g(\lambda-\mu) \,,
  \qquad
  \wh{f \star g} = \wh{f} \times \wh{g} \,.
\end{equation}

\paragraph{Properties of the functions $\phi_\alpha$ for $0<\alpha<\pi/2$ and $|\Im\,\lambda|<\alpha$:}
\begin{eqnarray} \label{eq:phidbl}
   \exp[i\phi_{2\alpha}(2\lambda)]&=&-\exp[i\phi_\alpha(\lambda) + i\phi_\alpha(\lambda+i\pi/2)] \,.\\
  \phi_\alpha(\pm\infty) &=& \pm (\pi-2\alpha) \,. \label{eq:phiinf} \\
 \phi'_\alpha(\lambda) &=& \frac{2\sin 2\alpha}{\cosh 2\lambda - \cos 2\alpha} \,.\\
  \wh{\phi'_\alpha}(\omega) &=&
  \frac{2\pi \sinh(\pi/2-\alpha)\omega}{\sinh \pi\omega/2} \,.
\end{eqnarray}

\paragraph{Computation of the quasi-momentum of a hole.} 
Using the $2i\pi$-periodicity of $k(\lambda)$ one can easily compute its Fourier transform by deforming contours
\begin{equation*}
k(\lambda):=\log \frac{\cosh\lambda +\sin\gamma}{\cosh\lambda -\sin\gamma}=\fint d\omega\,\frac{e^{-i\omega\lambda}\sinh\omega\gamma}{\omega\cosh\pi\omega/2}\ .
\end{equation*}
Notice that the  large $\lambda$ behaviour of the two hand sides agree. With this one gets
\begin{equation}
k_h(\lambda):=-[(1+J_\odd)\star k](\lambda)=\fint d\omega\, \frac{e^{-i\omega\lambda}}{2\omega \sinh (\pi/2-\gamma)\omega}\,.
\label{eq:tauh}
\end{equation}
To arrive at \eref{eq:sh} we use the periodicity of  $\sinh (\pi/2-\gamma)\omega$ to first compute $k_h'(\lambda)=\pi\tanh[\pi\lambda/(\pi-2\gamma)]/(\pi-2\gamma)$ and then integrate the result.
The integration constant can be fixed by comparing with the asymptotic of \eref{eq:tauh} at $\lambda\to\pm\infty$.

\paragraph{Dilogarithm integrals:}
\begin{equation} \label{eq:dilog1}
  \int_0^1 du \ \frac{\log(1+u)}{u} = \frac{\pi^2}{12}
\end{equation}
\begin{equation} \label{eq:dilog2}
  \Re\, \int_0^{\alpha} d(e^{i\theta}) \ \frac{\log(1+e^{i\theta})}{e^{i\theta}}
  = -\frac{\alpha^2}{4} \,,
  \qquad \text{for $-\pi<\alpha<\pi$.}
\end{equation}

\section*{References}

\bibliographystyle{unsrt}

\begin{thebibliography}{99}

\bibitem{Zamolodchikov:1978xm}
  A.~B.~Zamolodchikov and Al.~B.~Zamolodchikov,
  ``Factorized $S$-Matrices in Two-Dimensions as the Exact Solutions of Certain Relativistic Quantum Field Models,''
  Annals Phys.\  {\bf 120} (1979) 253.

\bibitem{Yang:1968rm}
  C.~-N.~Yang and C.~P.~Yang,
  ``Thermodynamics of one-dimensional system of bosons with repulsive delta function interaction,''
  J.\ Math.\ Phys.\  {\bf 10} (1969) 1115.

\bibitem{Zamolodchikov:1989cf}
  Al.~B.~Zamolodchikov,
  ``Thermodynamic Bethe Ansatz In Relativistic Models. Scaling Three State Potts And Lee-yang Models,''
  Nucl.\ Phys.\ B {\bf 342} (1990) 695.

\bibitem{Balog:2003xd}
  J.~Balog and A.~Hegedus,
  ``TBA equations for excited states in the sine-Gordon model,''
  J.\ Phys.\ A {\bf 37} (2004) 1903
  [hep-th/0304260].

\bibitem{Faddeev-Takh:1981}
  L.~D.~Faddeev and L.~A.~Takhtajan,
  ``What is the spin of a spin wave?,''
  Phys.\ Lett.\ A {\bf 85} (1981) 375.

\bibitem{Klummpe:1991vs}
  A.~Klumper, M.~T.~Batchelor and P.~A.~Pearce,
  ``Central charges of the 6- and 19- vertex models with twisted boundary conditions,''
  J.\ Phys.\ A {\bf 24} (1991) 3111.

\bibitem{Destri:1992qk}
  C.~Destri and H.~J.~de Vega,
  ``New thermodynamic Bethe ansatz equations without strings,''
  Phys.\ Rev.\ Lett.\  {\bf 69} (1992) 2313.


\bibitem{Destri:1994bv}
  C.~Destri and H.~J.~de Vega,
  ``Unified approach to thermodynamic Bethe Ansatz and finite size corrections for lattice models and field theories,''
  Nucl.\ Phys.\ B {\bf 438} (1995) 413
  [hep-th/9407117].


\bibitem{Destri:1997yz}
  C.~Destri and H.~J.~de Vega,
  ``Nonlinear integral equation and excited states scaling functions in the sine-Gordon model,''
  Nucl.\ Phys.\ B {\bf 504} (1997) 621
  [hep-th/9701107].

\bibitem{Feverati:1998dt}
  G.~Feverati, F.~Ravanini and G.~Takacs,
  ``Nonlinear integral equation and finite volume spectrum of Sine-Gordon theory,''
  Nucl.\ Phys.\ B {\bf 540} (1999) 543
  [hep-th/9805117].


\bibitem{Feverati:1998uz}
  G.~Feverati, F.~Ravanini and G.~Takacs,
  ``Scaling functions in the odd charge sector of sine-Gordon / massive Thirring theory,''
  Phys.\ Lett.\ B {\bf 444} (1998) 442
  [hep-th/9807160].



\bibitem{Hollowood:1992sy}
  T.~J.~Hollowood,
  ``Quantizing SL(N) solitons and the Hecke algebra,''
  Int.\ J.\ Mod.\ Phys.\ A {\bf 8} (1993) 947
  [hep-th/9203076].


\bibitem{ZinnJustin:1997at}
  P.~Zinn-Justin,
  ``Nonlinear integral equations for complex affine Toda models associated to simply laced Lie algebras,''
  J.\ Phys.\ A {\bf 31} (1998) 6747
  [hep-th/9712222].

\bibitem{Teschner:2007ng}
  J.~Teschner,
  ``On the spectrum of the Sinh-Gordon model in finite volume,''
  Nucl.\ Phys.\ B {\bf 799} (2008) 403
  [hep-th/0702214].

\bibitem{Arinshtein:1979pb}
  A.~E.~Arinshtein, V.~A.~Fateev and A.~B.~Zamolodchikov,
  ``Quantum $S$ Matrix of the (1+1)-Dimensional Todd Chain,''
  Phys.\ Lett.\ B {\bf 87} (1979) 389.
	
\bibitem{Braden:1989bu}
  H.~W.~Braden, E.~Corrigan, P.~E.~Dorey and R.~Sasaki,
  ``Affine Toda Field Theory and Exact S Matrices,''
  Nucl.\ Phys.\ B {\bf 338} (1990) 689.


\bibitem{Zamolodchikov:2000kt}
  Al.~B.~Zamolodchikov,
  ``On the thermodynamic Bethe ansatz equation in sinh-Gordon model,''
  J.\ Phys.\ A {\bf 39} (2006) 12863
  [hep-th/0005181].

\bibitem{Bytsko:2006ut}
  A.~G.~Bytsko and J.~Teschner,
  ``Quantization of models with non-compact quantum group symmetry: Modular XXZ magnet and lattice sinh-Gordon model,''
  J.\ Phys.\ A {\bf 39} (2006) 12927
  [hep-th/0602093].

\bibitem{Bytsko:2009mg}
  A.~Bytsko and J.~Teschner,
  ``The Integrable structure of nonrational conformal field theory,''
  arXiv:0902.4825 [hep-th].
	
\bibitem{Hikida:2008pe}
  Y.~Hikida and V.~Schomerus,
  ``The FZZ-Duality Conjecture: A Proof,''
  JHEP {\bf 0903} (2009) 095
  [arXiv:0805.3931 [hep-th]].
	
\bibitem{Witten:1991yr}
  E.~Witten,
  ``On string theory and black holes,''
  Phys.\ Rev.\ D {\bf 44} (1991) 314.
	
	
\bibitem{Jacobsen:2005xz}
  J.~L.~Jacobsen and H.~Saleur,
  ``The Antiferromagnetic transition for the square-lattice Potts model,''
  Nucl.\ Phys.\ B {\bf 743} (2006) 207
  [cond-mat/0512058].
	

\bibitem{Ikhlef:2008zz}
  Y.~Ikhlef, J.~Jacobsen and H.~Saleur,
  ``A staggered six-vertex model with non-compact continuum limit,''
  Nucl.\ Phys.\ B {\bf 789} (2008) 483.
	
\bibitem{Ikhlef:2011ay}
  Y.~Ikhlef, J.~L.~Jacobsen and H.~Saleur,
  ``An Integrable spin chain for the $\mathrm{SL}(2,\mathbb{R})/\mathrm{U}(1)$ black hole sigma model,''
  Phys.\ Rev.\ Lett.\  {\bf 108} (2012) 081601
  [arXiv:1109.1119 [hep-th]].
	

\bibitem{Hanany:2002ev}
  A.~Hanany, N.~Prezas and J.~Troost,
  ``The Partition function of the two-dimensional black hole conformal field theory,''
  JHEP {\bf 0204} (2002) 014
  [hep-th/0202129].
	
\bibitem{Schiff:1992tv}
  J.~Schiff,
  ``The Nonlinear Schrodinger equation and conserved quantities in the deformed parafermion and SL(2,R) / U(1) coset models,''
  hep-th/9210029.
	

\bibitem{Destri:1987ug}
  C.~Destri and H.~J.~de Vega,
  ``Light Cone Lattices And The Exact Solution Of Chiral Fermion And Sigma Models,''
  J.\ Phys.\ A {\bf 22} (1989) 1329.

\bibitem{Reshetikhin:1993wm}
  N.~Y.~Reshetikhin and H.~Saleur,
  ``Lattice regularization of massive and massless integrable field theories,''
  Nucl.\ Phys.\ B {\bf 419} (1994) 507
  [hep-th/9309135].
		
	\bibitem{IKB}
V.~E.~Korepin, N.~M.~Bogoliubov and A.~G.~Izergin,  ``Quantum inverse scattering method and correlation functions.'' (1997) Cambridge university press.

\bibitem{Alcaraz:1987zr}
  F.~C.~Alcaraz, M.~N.~Barber and M.~T.~Batchelor,
  ``Conformal Invariance, The XXZ Chain And The Operator Content Of Two-dimensional Critical Systems,''
  Annals Phys.\  {\bf 182} (1988) 280.

\bibitem{Lukyanov}
S.~L.~Lukyanov, unpublished notes.


\bibitem{Yang:1966sa}
  C.~N.~Yang and C.~P.~Yang,
  ``One-dimensional chain of anisotropic spin spin interactions. 2. Properties of the ground state energy per lattice site for an infinite system,''
  Phys.\ Rev.\  {\bf 150} (1966) 327.


\bibitem{Fateev:1990bf}
  V.~A.~Fateev,
  ``Integrable deformations in Z(N) symmetrical models of conformal quantum field theory,''
  Int.\ J.\ Mod.\ Phys.\ A {\bf 6} (1991) 2109.

\bibitem{Bakas:1991fs}
  I.~Bakas and E.~Kiritsis,
  ``Beyond the large N limit: Nonlinear W(infinity) as symmetry of the SL(2,R) / U(1) coset model,''
  Int.\ J.\ Mod.\ Phys.\ A {\bf 7S1A} (1992) 55
   [Int.\ J.\ Mod.\ Phys.\ A {\bf 7} (1992) 55]
  [hep-th/9109029].

\bibitem{Yu:1991cv}
  F.~Yu and Y.~-S.~Wu,
  ``Nonlinear W(hat)(infinity) current algebra in the SL(2,R) / U(1) coset model,''
  Phys.\ Rev.\ Lett.\  {\bf 68} (1992) 2996
   [Erratum-ibid.\  {\bf 69} (1992) 554]
  [hep-th/9112009].
	
	
	









\bibitem{Bazhanov:1994ft}
  V.~V.~Bazhanov, S.~L.~Lukyanov and A.~B.~Zamolodchikov,
  ``Integrable structure of conformal field theory, quantum KdV theory and thermodynamic Bethe ansatz,''
  Commun.\ Math.\ Phys.\  {\bf 177} (1996) 381
  [hep-th/9412229].
	
\bibitem{Bazhanov:1996dr}
  V.~V.~Bazhanov, S.~L.~Lukyanov and A.~B.~Zamolodchikov,
  ``Integrable structure of conformal field theory. 2. Q operator and DDV equation,''
  Commun.\ Math.\ Phys.\  {\bf 190} (1997) 247
  [hep-th/9604044].
	
\bibitem{Bazhanov:1998dq}
  V.~V.~Bazhanov, S.~L.~Lukyanov and A.~B.~Zamolodchikov,
  ``Integrable structure of conformal field theory. 3. The Yang-Baxter relation,''
  Commun.\ Math.\ Phys.\  {\bf 200} (1999) 297
  [hep-th/9805008].
	
\bibitem{Dorey:2007zx}
  P.~Dorey, C.~Dunning and R.~Tateo,
  ``The ODE/IM Correspondence,''
  J.\ Phys.\ A {\bf 40} (2007) R205
  [hep-th/0703066].

	
\bibitem{Fateev:2005kx}
  V.~A.~Fateev and S.~L.~Lukyanov,
  ``Boundary RG flow associated with the AKNS soliton hierarchy,''
  J.\ Phys.\ A {\bf 39} (2006) 12889
  [hep-th/0510271].
	

\bibitem{Lukyanov:2013wra}
  S.~L.~Lukyanov,
  ``ODE/IM correspondence for the Fateev model,''
  arXiv:1303.2566 [hep-th].

	
\bibitem{Pohlmeyer:1975nb}
  K.~Pohlmeyer,
  ``Integrable Hamiltonian Systems and Interactions Through Quadratic Constraints,''
  Commun.\ Math.\ Phys.\  {\bf 46} (1976) 207.
	
\bibitem{Lund:1976ze}
  F.~Lund and T.~Regge,
  ``Unified Approach to Strings and Vortices with Soliton Solutions,''
  Phys.\ Rev.\ D {\bf 14} (1976) 1524.
	
\bibitem{Sciuto:1979ta}
  S.~Sciuto,
  ``Exterior Calculus And Two-dimensional Supersymmetric Models,''
  Phys.\ Lett.\ B {\bf 90} (1980) 75.
	
\bibitem{Getmanov:1980cq}
  B.~S.~Getmanov,
  ``Integrable Two-dimensional Lorentz Invariant Nonlinear Model Of Complex Scalar Field (complex Sine-gordon II),''
  Theor.\ Math.\ Phys.\  {\bf 48} (1982) 572
   [Teor.\ Mat.\ Fiz.\  {\bf 48} (1981) 13].
	
\bibitem{Fateev:1996em}
  V.~A.~Fateev,
  ``Integrable deformations of affine Toda theories and duality,''
  Nucl.\ Phys.\ B {\bf 479} (1996) 594.
	
\bibitem{deVega:1981ka}
  H.~J.~de Vega and J.~M.~Maillet,
  ``Renormalization Character And Quantum S Matrix For A Classically Integrable Theory,''
  Phys.\ Lett.\ B {\bf 101} (1981) 302.
	
\bibitem{deVega:1982sh}
  H.~J.~de Vega and J.~M.~Maillet,
  ``Semiclassical Quantization Of The Complex Sine-gordon Field Theory,''
  Phys.\ Rev.\ D {\bf 28} (1983) 1441.
	
	
\bibitem{Bonneau:1984pj}
  G.~Bonneau and F.~Delduc,
  ``S Matrix Properties Versus Renormalizability In Two-dimensional O(n) Symmetric Models,''
  Nucl.\ Phys.\ B {\bf 250} (1985) 561.
	
\bibitem{Fateev:1995ht}
  V.~A.~Fateev,
  ``The Duality between two-dimensional integrable field theories and sigma models,''
  Phys.\ Lett.\ B {\bf 357} (1995) 397.
	
\bibitem{Ridout:2011wx}
  D.~Ridout and J.~Teschner,
  ``Integrability of a family of quantum field theories related to sigma models,''
  Nucl.\ Phys.\ B {\bf 853} (2011) 327
  [arXiv:1102.5716 [hep-th]].
	
\bibitem{Dorey:1994mg}
  N.~Dorey and T.~J.~Hollowood,
  ``Quantum scattering of charged solitons in the complex sine-Gordon model,''
  Nucl.\ Phys.\ B {\bf 440} (1995) 215
  [hep-th/9410140].
	
	
	\bibitem{HS}
	H.~Saleur, unpublished notes.
	
\bibitem{Zirnbauer:1999}
  M.~R.~Zirnbauer,
  ``Conformal Field Theory of the Integer Quantum Hall Plateau Transition,''
  {[hep-th/9905054]}.

\bibitem{GLR:1999}
  I.~Gruzberg, A.~W.~W.~Ludwig and N.~Read,
  ``Exact exponents for the spin quantum Hall transition,''
  Phys.\ Rev.\ Lett.\ {\bf 82} (1999) 4524
  {[cond-mat/9902063]}

\bibitem{Essler:2005ag}
  F.~H.~L.~Essler, H.~Frahm and H.~Saleur,
  ``Continuum limit of the integrable sl(2/1) 3 - anti-3 superspin chain,''
  Nucl.\ Phys.\ B {\bf 712} (2005) 513
  [cond-mat/0501197].
	
\bibitem{Candu:2010qj}
  C.~Candu,
  ``Continuum Limit of gl(M/N) Spin Chains,''
  JHEP {\bf 1107} (2011) 069
  [arXiv:1012.0050 [hep-th]].
	
\bibitem{Frahm:2010ei}
  H.~Frahm and M.~J.~Martins,
  ``Finite size properties of staggered $U_q[sl(2|1)]$ superspin chains,''
  Nucl.\ Phys.\ B {\bf 847} (2011) 220
  [arXiv:1012.1753 [cond-mat.stat-mech]].
	
\bibitem{Frahm:2012eb}
  H.~Frahm and M.~J.~Martins,
  ``Phase Diagram of an Integrable Alternating $U_q[sl(2|1)]$ Superspin Chain,''
  Nucl.\ Phys.\ B {\bf 862} (2012) 504
  [arXiv:1202.4676 [cond-mat.stat-mech]].
	
	
\end{thebibliography}

\end{document}